%% file: main.tex
\documentclass[twocolumn]{aastex631}
\input{affil}
\usepackage{CJK}
\usepackage[caption=false]{subfig}

\def\EE#1{\times 10^{#1}}
\def\kms{\mbox{\,km~s$^{-1}$}}
\def\msun{\hbox{\,$M_{\odot}$}}
\def\msunyr{\mbox{\,$M_\odot{\rm \, yr^{-1}}$}}
\def\cm{\mbox{\,cm}}
\def\cm3{\mbox{\,cm$^{-3}$}}

\shorttitle{SN~2021aefx}
\shortauthors{Hosseinzadeh et al.}
\submitjournal{\textit{The Astrophysical Journal Letters}}
\received{2022 May 4}
\revised{2022 June 9}
\accepted{2022 June 27}

\begin{document}

\title{Constraining the Progenitor System of the Type~Ia Supernova 2021aefx}

\correspondingauthor{Griffin Hosseinzadeh}
\email{griffin0@arizona.edu}

\author[0000-0002-0832-2974]{Griffin Hosseinzadeh}
\UA
\author[0000-0003-4102-380X]{David J.\ Sand}
\UA
\author[0000-0002-3664-8082]{Peter Lundqvist}
\OKC

\author[0000-0003-0123-0062]{Jennifer E.\ Andrews}
\GeminiNorth
\author[0000-0002-4924-444X]{K.\ Azalee Bostroem}
\DiRAC\UW
\author[0000-0002-7937-6371]{Yize Dong \begin{CJK*}{UTF8}{gbsn}(董一泽)\end{CJK*}}
\UCD
\author[0000-0003-0549-3281]{Daryl Janzen}
\USask
\author[0000-0001-5754-4007]{Jacob E.\ Jencson}
\UA
\author[0000-0001-9589-3793]{Michael Lundquist}
\Keck
\author[0000-0002-7015-3446]{Nicolas E.\ Meza Retamal}
\UCD
\author[0000-0002-0744-0047]{Jeniveve Pearson}
\UA
\author[0000-0001-8818-0795]{Stefano Valenti}
\UCD
\author[0000-0003-2732-4956]{Samuel Wyatt}
\UA

\author[0000-0003-0035-6659]{Jamison Burke}
\LCO\UCSB
\author[0000-0003-4253-656X]{D.\ Andrew Howell}
\LCO\UCSB
\author[0000-0001-5807-7893]{Curtis McCully}
\LCO\UCSB
\author{Megan Newsome}
\LCO\UCSB
\author[0000-0003-0209-9246]{Estefania Padilla Gonzalez}
\LCO\UCSB
\author[0000-0002-7472-1279]{Craig Pellegrino}
\LCO\UCSB
\author[0000-0003-0794-5982]{Giacomo Terreran}
\LCO\UCSB

\author[0000-0003-3108-1328]{Lindsey A.\ Kwok}
\Rutgers
\author[0000-0001-8738-6011]{Saurabh W.\ Jha}
\Rutgers
\author[0000-0002-1468-9668]{Jay Strader}
\MSU
\author[0000-0002-4807-379X]{Esha Kundu}
\MSU
\author[0000-0003-4501-8100]{Stuart D.\ Ryder}
\Macquarie\AAARC

\author[0000-0002-6703-805X]{Joshua Haislip}
\UNC
\author[0000-0003-3642-5484]{Vladimir Kouprianov}
\UNC
\author[0000-0002-5060-3673]{Daniel E.\ Reichart}
\UNC

\begin{abstract}

We present high-cadence optical and ultraviolet light curves of the normal Type~Ia supernova (SN) 2021aefx, which shows an early bump during the first two days of observation. This bump may be a signature of interaction between the exploding white dwarf and a nondegenerate binary companion, or it may be intrinsic to the white dwarf explosion mechanism. In the case of the former, the short duration of the bump implies a relatively compact main-sequence companion star, although this conclusion is viewing-angle dependent. Our best-fit companion-shocking and double-detonation models both overpredict the UV luminosity during the bump, and existing nickel-shell models do not match the strength and timescale of the bump. We also present nebular spectra of SN~2021aefx, which do not show the hydrogen or helium emission expected from a nondegenerate companion, as well as a radio nondetection that rules out all symbiotic progenitor systems and most accretion disk winds. Our analysis places strong but conflicting constraints on the progenitor of SN~2021aefx; no current model can explain all of our observations.

\end{abstract}


\section{Introduction\label{sec:intro}}
The explosions of carbon-oxygen white dwarfs (WDs) as Type~Ia supernovae (SNe~Ia) have long been important tools in astrophysics, their use as standardizable candles having led to the discovery of dark energy \citep{riess_observational_1998,perlmutter_measurements_1999}. However, despite decades of intense study, SN~Ia progenitor systems and explosion mechanisms are still not fully understood (see reviews by \citealt{howell_type_2011}, \citealt{maoz_observational_2014}, \citealt{maguire_type_2017}, and \citealt{jha_observational_2019}). The progenitor WD must be in a binary (or higher) system in order to accrete material and ignite, but the companion star may either be a main-sequence or giant star \citep[the single-degenerate scenario;][]{whelan_binaries_1973} or another WD \citep[the double-degenerate scenario;][]{iben_supernovae_1984,webbink_double_1984}. Additionally, the explosion may be a pure deflagration \citep{nomoto_accreting_1984}, a deflagration that transitions to a detonation after the WD is unbound \citep[a delayed detonation;][]{khokhlov_delayed_1991}, a detonation triggered by an earlier helium-shell detonation \citep[a double detonation;][]{nomoto_accreting_1982,livne_successive_1990,woosley_sub-chandrasekhar_2011}, or another more extreme model \citep[e.g., a dynamical or violent merger;][]{benz_three-dimensional_1989,loren-aguilar_high-resolution_2009,katz_rate_2012,kushnir_head-on_2013}. See \cite{hoeflich_explosion_2017} for a review of explosion mechanisms.

The very early light curves of SNe~Ia are among the best probes of the progenitor system and explosion physics. Whereas most of their evolution is uniform (such that they can be used for cosmology), their light curves during the first few days after explosion can reveal important clues about their progenitor systems and explosion triggers. For example, \cite{kasen_seeing_2010} predicted that, given a favorable viewing angle, the collision between the SN ejecta and a nondegenerate companion star would result in an observable UV excess over a smooth rise, an effect that would disappear after only a few days (when most SNe are discovered). However, such a ``bump'' might also be explained by an unusual distribution of radioactive $^{56}$Ni in the SN ejecta \citep[e.g.,][]{magee_investigation_2020}, or it might be an expected feature of a sub-Chandrasekhar-mass explosion \citep[e.g.,][]{polin_observational_2019}.

Over the past decade, only a handful of nearby SNe~Ia have been discovered early enough to test these hypotheses. SN~2011fe followed a smooth $L \propto t^2$ rise, from which \cite{nugent_supernova_2011} and \cite{bloom_compact_2012} derived a stellar radius that confirmed a WD progenitor. \cite{li_exclusion_2011} and \cite{nugent_supernova_2011} also rule out a nondegenerate companion to SN~2011fe at high significance.
\cite{cao_strong_2015} showed an initial decline in the UV light curve of the peculiar (02es-like) SN~Ia iPTF14atg, and
\cite{marion_sn_2016} showed a slight excess in the early $U$ and $B$ light curves of the normal Type~Ia SN~2012cg compared to a power-law rise, both of which were interpreted as interaction with a nondegenerate companion (though see \citealt{shappee_strong_2018} for the latter event).
\cite{hosseinzadeh_early_2017} presented a time-resolved excess in the early $U$, $B$, and $g$ light curves of SN~2017cbv, with a ${\sim}6$-hour cadence over the first 5 days, which could be explained by a nondegenerate companion. However, the companion-shocking model greatly overpredicted the strength of the bump in the UV bands.
\cite{jiang_hybrid_2017} concluded that the early, red flash in MUSSES1604D arose from a helium-shell detonation.
\cite{miller_early_2018} presented a ${\sim}2$~mag rise over the first day of the $g$-band light curve of iPTF16abc, which they interpreted as the effect of circumstellar interaction and/or nickel mixing.
\cite{dimitriadis_k2_2019} and \cite{shappee_seeing_2019} both presented a two-component power-law rise in the high-cadence Kepler~2 light curve of SN~2018oh \citep{li_photometric_2019}, but they disagree about its origin with respect to the companion-shocking, nickel-mixing, and double-detonation models.
\cite{miller_spectacular_2020}, \cite{burke_bright_2021}, and \cite{tucker_sn_2021} showed that the extremely luminous UV excess during the first ${\sim}5$ days of the 02es-like SN~2019yvq also could not be fully explained by any single model, although \cite{siebert_strong_2020} conclude it was a double detonation.
\cite{jiang_discovery_2021} concluded that the ${<}1$~day flash in SN~2020hvf was due to circumstellar interaction.
\cite{ni_infant-phase_2022} presented a unique red bump in the early light curve of SN~2018aoz, which they interpreted as an overdensity of iron-peak elements in the outer layers of the ejecta.
Most recently, \cite{deckers_constraining_2022} found six new events with early excesses in $g$ and/or $r$ band in a sample of 115 SNe~Ia, and J.~Burke et al.\ (2022, in preparation) will present a uniform analysis of 9 nearby well-observed SNe~Ia.

An additional probe of the nature of the binary companion comes from nebular spectroscopy. In the case of a nondegenerate companion, the SN ejecta should be enriched with hydrogen and/or helium stripped from the companion, which should appear in emission once the ejecta are mostly transparent \citep{marietta_type_2000,pan_impact_2010,pan_impact_2012,liu_three-dimensional_2012,liu_impact_2013,botyanszki_multidimensional_2018,dessart_spectral_2020}. \cite{lundqvist_hydrogen_2013,lundqvist_no_2015} note that forbidden oxygen and calcium lines may also arise from interaction with a nondegenerate companion, but these lines are weak and difficult to disentangle from the underlying SN emission.

\begin{figure*}
    \centering
    \includegraphics[width=\textwidth]{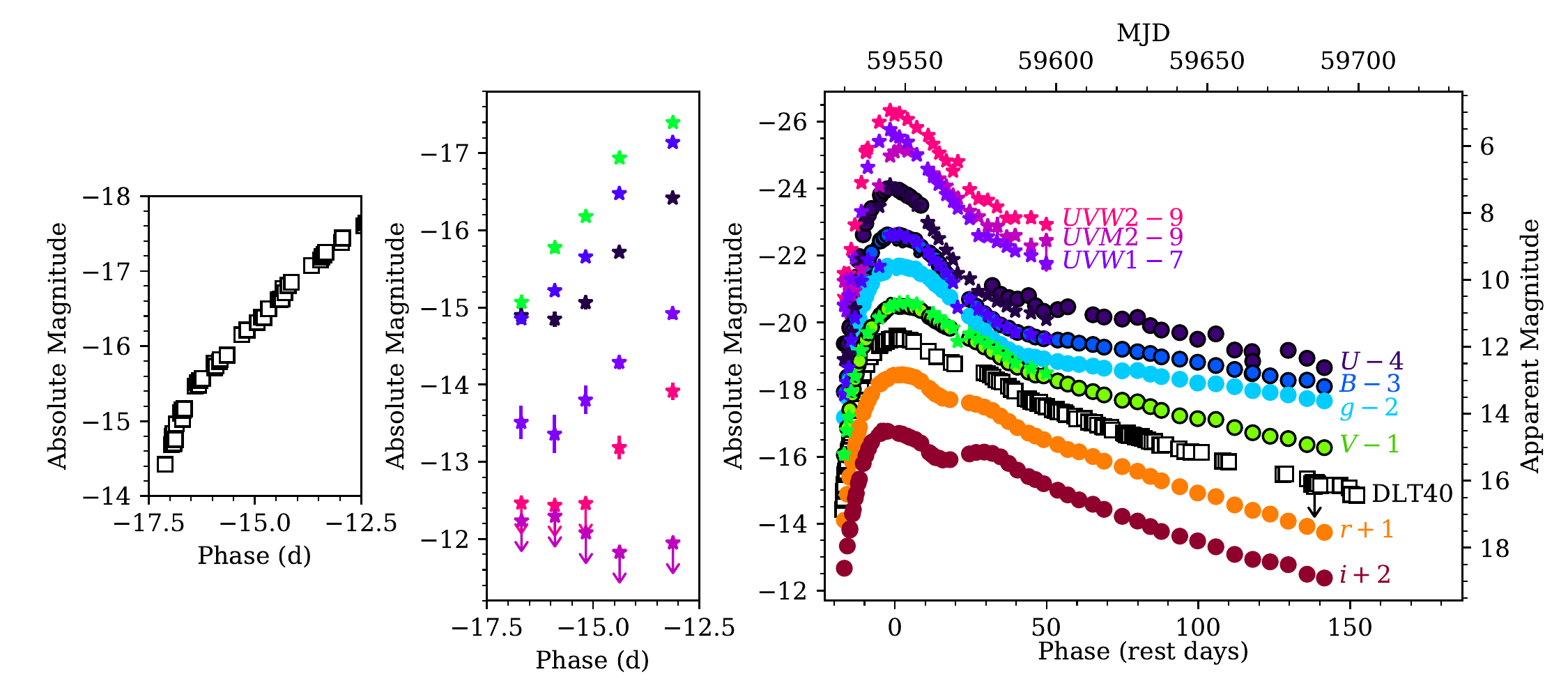}
    \caption{Light curve of SN~2021aefx from Las Cumbres Observatory (circles), DLT40 (squares), and Swift (stars), offset and binned by 0.01~days for clarity. Phase is given with respect to maximum light in the $B$ band, estimated in \S\ref{sec:phot}. The left panel highlights the very high-cadence unfiltered light curve from DLT40, with no offset, showing a bump lasting ${\sim}2$ days. The middle panel highlights the early Swift photometry, with no offsets, showing an initial decrease in the $U$ and $UVW1$ filters.\\(The data used to create this figure are available.)}
    \label{fig:lc}
\end{figure*}

Hydrogen emission has been detected in the so-called SNe~Ia-CSM, which are thought to have symbiotic nova progenitors \citep[e.g.][]{dilday_ptf_2012}, but these are only a rare subclass of SNe~Ia \citep[${\lesssim}5\%$;][]{dubay_late-onset_2022}. Narrow emission lines have not been detected in the vast majority of normal SNe~Ia (e.g., \citealt{maguire_searching_2016,sand_nebular_2019,tucker_nebular_2020} for recent compilations), even where the light curve showed an early bump \citep[e.g.,][]{sand_nebular_2018,dimitriadis_nebular_2019,tucker_no_2019}. This appears to disfavor the single-degenerate scenario for most normal SNe~Ia. However, three recent detections of H$\alpha$ in fast-declining SNe~Ia suggest that this channel may be physically allowed for some subluminous SNe~Ia \citep{kollmeier_h_2019,vallely_asassn-18tb:_2019,prieto_variable_2020,elias-rosa_nebular_2021}.

Here we present the case of the normal Type~Ia SN~2021aefx, which was discovered within hours of explosion by the Distance Less Than 40~Mpc (DLT40) survey \citep{tartaglia_early_2018}. The early unfiltered light curve from DLT40, with an average cadence of ${\approx}2$ hr over the first 5 days, shows a subtle excess with respect to a power-law rise during the first two days of observation. In addition, its daily-cadence UV light curve shows an initial decrease over the first two observations. We refer to these and similar excesses in the $U$, $B$, and $g$ bands as the ``bump'' throughout our analysis. After describing the observations of SN~2021aefx in \S\ref{sec:obs}, we compare the light curve to models of companion shocking, sub-Chandrasekhar-mass WD explosions, and explosions of WDs with surface nickel in \S\ref{sec:phot}, in an attempt to explain the cause of the early bump. In \S\ref{sec:spec_early}, we describe the unique first spectrum of SN~2021aefx, and in \S\ref{sec:nebular} we search for hydrogen and helium emission in its nebular spectra. In \S\ref{sec:radio} we present constraints from the nondetection of SN~2021aefx in the radio. We discuss the implications of this analysis for its progenitor system and explosion mechanism in \S\ref{sec:discuss}.

\vspace{-1pt}
\section{Observations and Data Reduction\label{sec:obs}}
SN~2021aefx was discovered by DLT40 on 2021-11-11.343 UT at an unfiltered brightness of $17.25 \pm 0.05$~mag and was not detected on 2021-11-06.328 UT to an unfiltered limit of 19.35~mag \citep{valenti_dlt40_2021}.
Its J2000 coordinates are $\alpha=04\textsuperscript{h}19\textsuperscript{m}53\fs400$ and $\delta=-54\degr56'53\farcs09$ according to Gaia Photometric Science Alerts,\footnote{\url{https://www.wis-tns.org/object/2021aefx}} $1\farcm2$ (6.2~kpc) southeast of the center of its intermediate spiral host, NGC~1566 \citep{devaucouleurs_vizier_1995}. 
It was initially classified as a Type~I (hydrogen-poor) SN by \cite{bostroem_global_2021} based on a unique early spectrum (see \S\ref{sec:spec_early} for more details).
Its Type~Ia classification was secured the next day by \cite{onori_epessto+_2021}.

\defcitealias{sdsscollaboration_thirteenth_2017}{SDSS Collaboration (2017)}

\subsection{UV/Optical Data}
We immediately initiated a high-cadence photometric follow-up campaign using the 1\,m and 0.4\,m telescope networks of Las Cumbres Observatory \citep{brown_cumbres_2013} and the Ultraviolet/Optical Telescope \citep[UVOT;][]{roming_swift_2005} on the Neil Gehrels Swift Observatory \citep{gehrels_swift_2004}.
We measured aperture photometry on the Las Cumbres images using \texttt{lcogtsnpipe} \citep{valenti_diversity_2016} and calibrated to images of the RU149 standard field \citep{rubin_second_1974} taken on the same night with the same telescope.
RU149 catalogs are from \cite{stetson_homogeneous_2000} for \textit{UBV} (Vega magnitudes) and the \citetalias{sdsscollaboration_thirteenth_2017} for \textit{gri} (AB magnitudes).
We measured photometry on the UVOT images using the High-Energy Astrophysics Software \citep[HEASoft;][]{nasaheasarc_heasoft:_2014} in a 3$\arcsec$ aperture centered at the position of SN\,2021aefx. We used a set of 10 UVOT pre-explosion archival observations taken between 2020 June 12 and 2020 October 21 to subtract the background contamination by emission from the host galaxy in the same aperture. We applied the zero-points of \cite{breeveld_updated_2011} with the time-dependent sensitivity corrections updated in 2020.
We correct for Milky Way extinction of $E(B-V)=0.0079$~mag \citep{schlafly_measuring_2011} and host-galaxy extinction of $E(B-V)=0.097$~mag (see \S\ref{sec:extinction}), according to the \cite{fitzpatrick_correcting_1999} extinction law, and a \cite{tully_new_1977} distance of $18.0 \pm 2.0$~Mpc \citep[$\mu=31.28 \pm 0.23$~mag;][]{sabbi_resolved_2018}.
The resulting light curve of SN~2021aefx is shown in Fig.~\ref{fig:lc}.

We also obtained optical spectroscopy of SN~2021aefx using the FLOYDS spectrograph on Las Cumbres Observatory's 2\,m Faulkes Telescope South \citep[FTS;][]{brown_cumbres_2013}, the High Resolution Spectrograph (HRS) and the Robert Stobie Spectrograph (RSS) on the South African Large Telescope \citep[SALT;][]{buckley_status_2006}, and the Goodman High-Throughput Spectrograph on the Southern Astrophysical Research Telescope \citep[SOAR;][]{clemens_goodman_2004}.
We reduced these spectra using the FLOYDS pipeline \citep{valenti_first_2014}, the SALT HRS MIDAS pipeline \citep{kniazev_mn48:_2016,kniazev_salt_2017}, PySALT \citep{crawford_pysalt:_2010}, the Goodman pipeline \citep{torres_goodman_2017}, and/or other standard PyRAF routines \citep{sciencesoftwarebranchatstsci_pyraf:_2012}. We also include the public classification spectrum of \cite{onori_epessto+_2021} from the ESO Faint Object Spectrograph and Camera 2 (EFOSC2) on the New Technology Telescope \citep[NTT;][]{buzzoni_eso_1984}.
The spectral series is logged in Table~\ref{tab:spec} and plotted in Fig.~\ref{fig:spec}, corrected for the redshift of the host galaxy \citep[$z=0.005017$;][]{allison_search_2014}.

\input{specplot_list}\vspace{-24pt}

\begin{figure*}
    \centering
    \includegraphics[width=\textwidth]{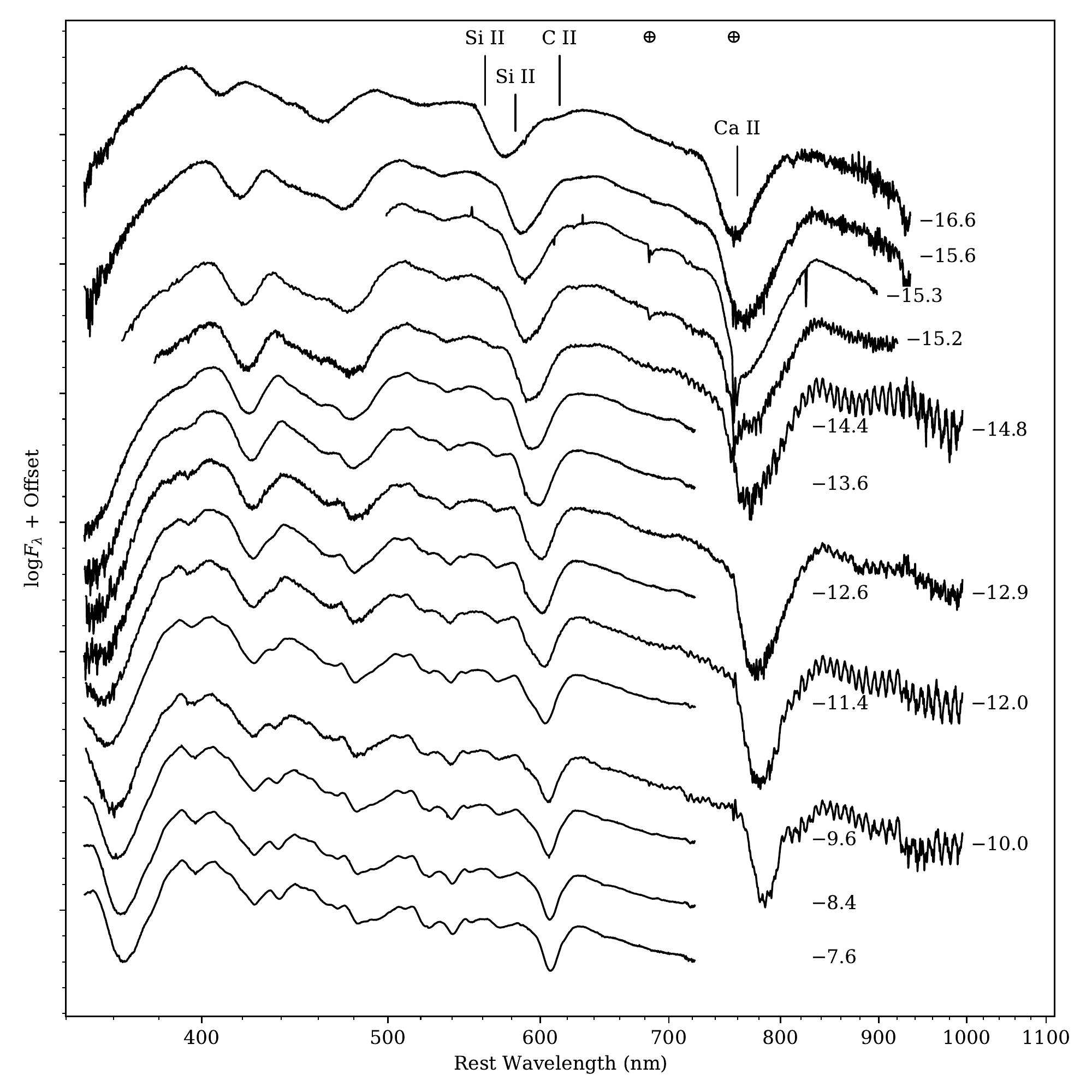}
    \caption{Low-resolution spectra of SN~2021aefx. Phases are marked to the right of each spectrum in rest-frame days after the $B$-band peak. The strongest absorption features in the first spectrum are marked. See Fig.~\ref{fig:spec_early} for a more detailed view. Telluric wavelengths are marked with the $\earth$ symbol.\\(The data used to create this figure are available.)}
    \label{fig:spec}
\end{figure*}

\begin{figure*}
    \centering
    \ContinuedFloat
    \includegraphics[width=\textwidth]{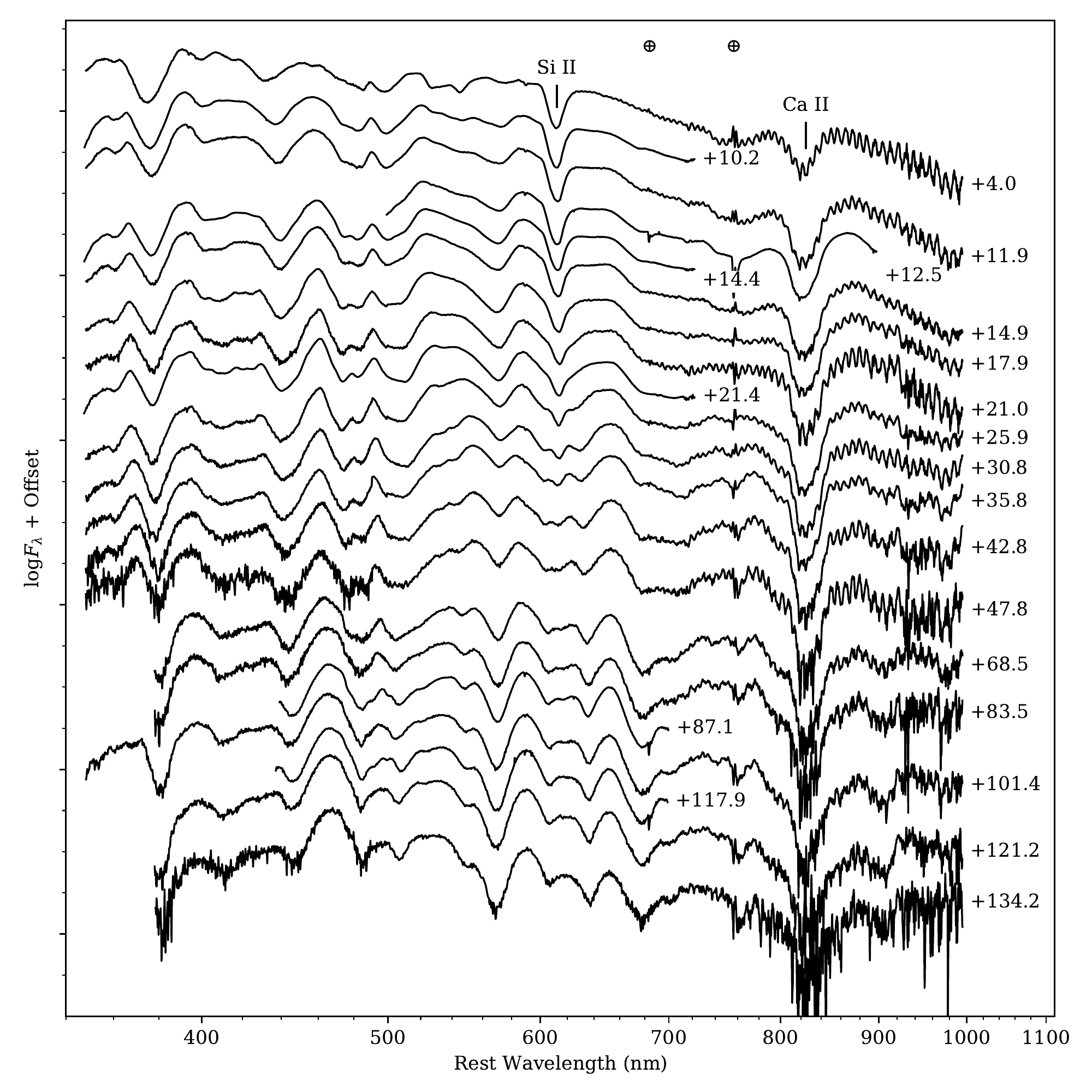}
    \caption{Continued.}
\end{figure*}

\subsection{Radio Data}
SN~2021aefx was observed in the radio with the Australia Telescope Compact Array (ATCA)
on two occasions \citep{kundu_mass-loss_2021}: 2021 November 13 between UT 14:30 and 17:30, and 2021 November 19 from UT 13:00 to 17:00. 
For the first epoch, the central frequencies were 5.5 GHz, 9 GHz, and 18 GHz, using a bandwidth of 2 GHz, while for the second epoch only the central frequencies 5.5 GHz and 9 GHz were used.
The observation and reduction procedures follow \citet{lundqvist_deepest_2020}. PKS~B1934$-$638 was used as the primary calibrator, and the quasar J0441$-$5154 was used as the secondary calibrator.

As reported by \cite{kundu_mass-loss_2021}, no radio emission was detected down to 3$\sigma$ upper limits of 70 $\mu$Jy \,beam$^{-1}$ at 5.5 GHz,
60 $\mu$Jy \,beam$^{-1}$ at 9.0 GHz, and 50 $\mu$Jy\,beam$^{-1}$ at 18 GHz for the first epoch, and 40 $\mu$Jy \,beam$^{-1}$ at 5.5 GHz and
30 $\mu$Jy \,beam$^{-1}$ at 9.0 GHz for the second. The total on-source time at each frequency was 1.0~hr during the first epoch and 3.0~hr
during the second. For a host-galaxy distance of 18.0~Mpc, this implies upper limits on the luminosity at the first epoch of 
$2.71~(2.33)~(1.94)\EE{25}$ erg s$^{-1}$ Hz$^{-1}$ for 5.5~(9)~(18) GHz and $1.55~(1.16)\EE{25}$ erg s$^{-1}$ Hz$^{-1}$ for 5.5~(9) GHz
at the second. 

\subsection{Host-galaxy Extinction\label{sec:extinction}}
\begin{figure}
    \centering
    \includegraphics[width=\columnwidth]{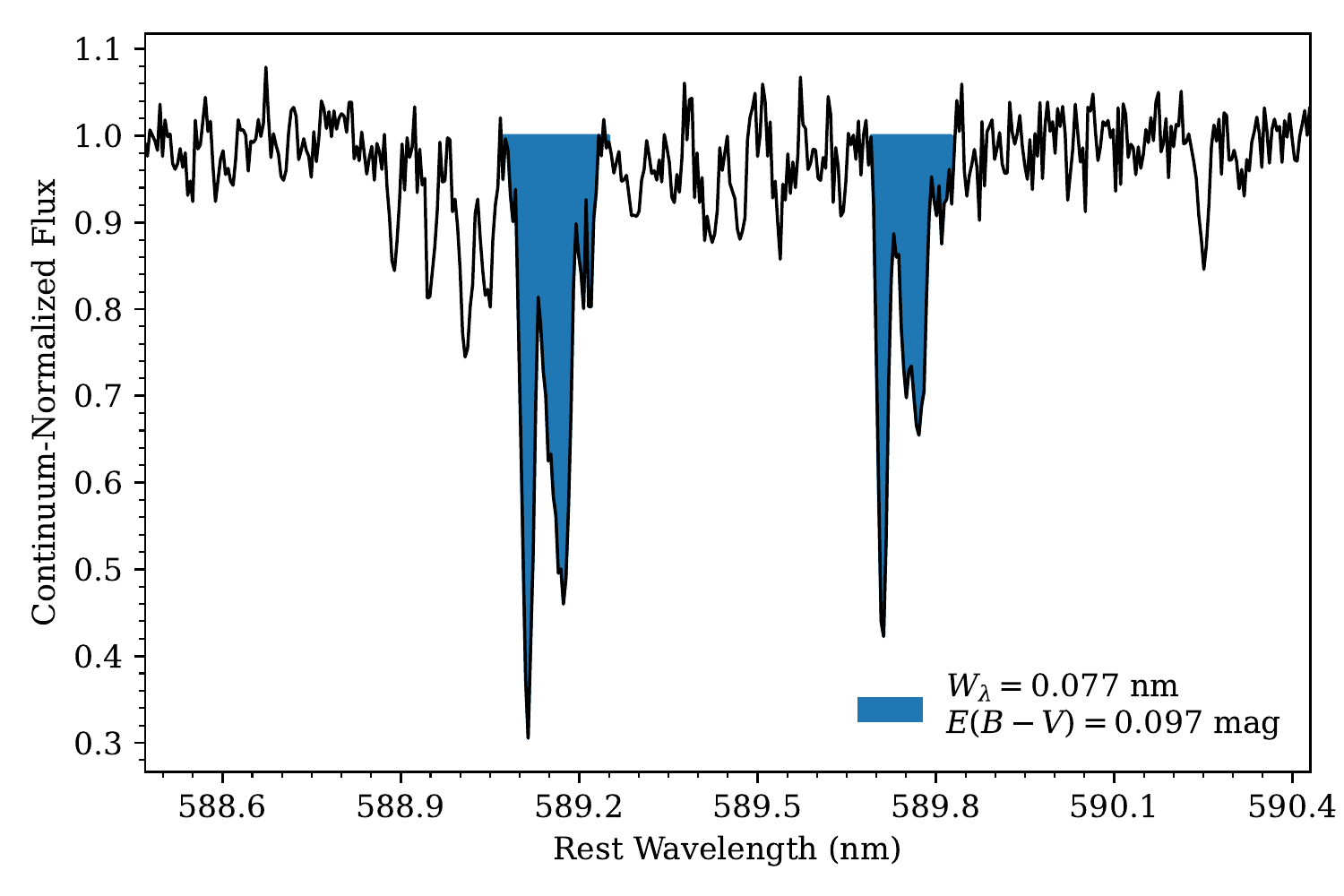}
    \includegraphics[width=\columnwidth]{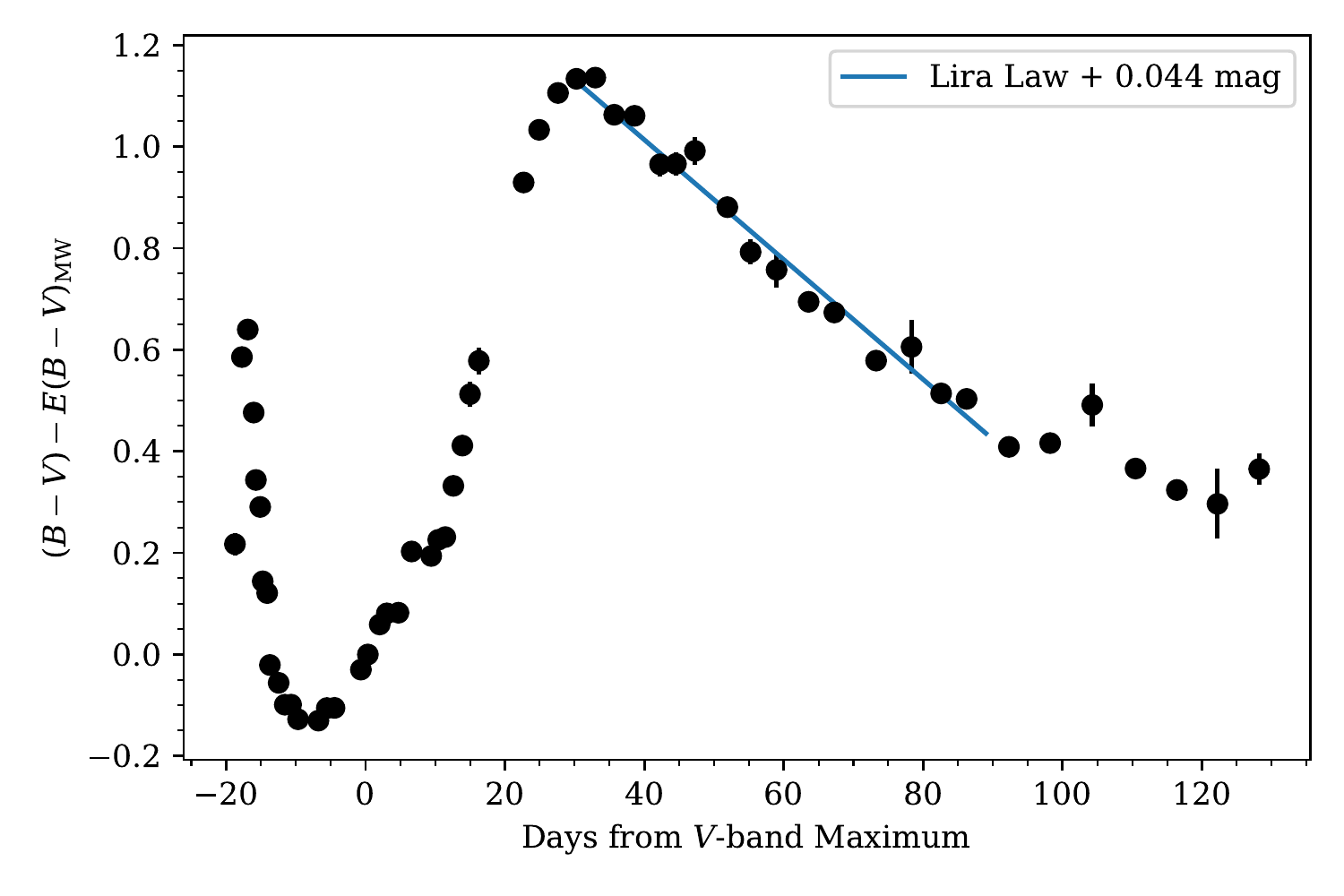}
    \caption{Top: high-resolution spectrum of SN~2021aefx in the region around \ion{Na}{1D}. We measure an equivalent width of $W_\lambda = 0.077$~nm, corresponding to a host-galaxy extinction of $E(B-V)_\mathrm{host}=0.097$~mag.
    Bottom: Milky Way extinction-corrected $B-V$ color curve of SN~2021aefx. A linear fit to the color on the radioactively powered tail shows an offset of $E(B-V)_\mathrm{host}=0.044$~mag with respect to the Lira Law. These are both consistent to within the scatter in the Lira Law, so we adopt the former estimate: $E(B-V)_\mathrm{host}=0.097$~mag.}
    \label{fig:extinction}
\end{figure}

Our high-resolution ($R \approx 40,000$) SALT spectrum, taken 16.4 days after peak, shows strong \ion{Na}{1D} in absorption at the redshift of the host galaxy (Fig.~\ref{fig:extinction}, top). We measure an equivalent width of $W_\lambda = 0.077$~nm, which corresponds to a host-galaxy extinction of $E(B-V)=0.097$~mag, according to the relationship of \cite{poznanski_empirical_2012} with the recalibration of \cite{schlafly_blue_2010}.

We also estimate the host-galaxy extinction by comparing to the \cite{lira_light_1996} Law, following the procedure of \cite{phillips_reddening-free_1999}. We fit a straight line with a fixed slope of $-0.0118$~mag~day$^{-1}$ to the $B-V$ color curve of SN~2021aefx between 30 and 60 days after $V$-band maximum, corrected for Milky Way extinction only, using weighted least-squares fitting (Fig.~\ref{fig:extinction}, bottom). This line is offset from the Lira Law by $E(B-V)_\mathrm{host}=0.044$~mag. Given that our two estimates agree to within the scatter in the Lira Law \citep[${\sim}0.05$ mag;][]{phillips_reddening-free_1999}, we adopt the former estimate, $E(B-V)_\mathrm{host}=0.097$~mag, throughout the remainder of our analysis.

\section{Light Curves and Color Curves\label{sec:phot}}
\begin{figure*}
    \centering
    \includegraphics[width=\textwidth]{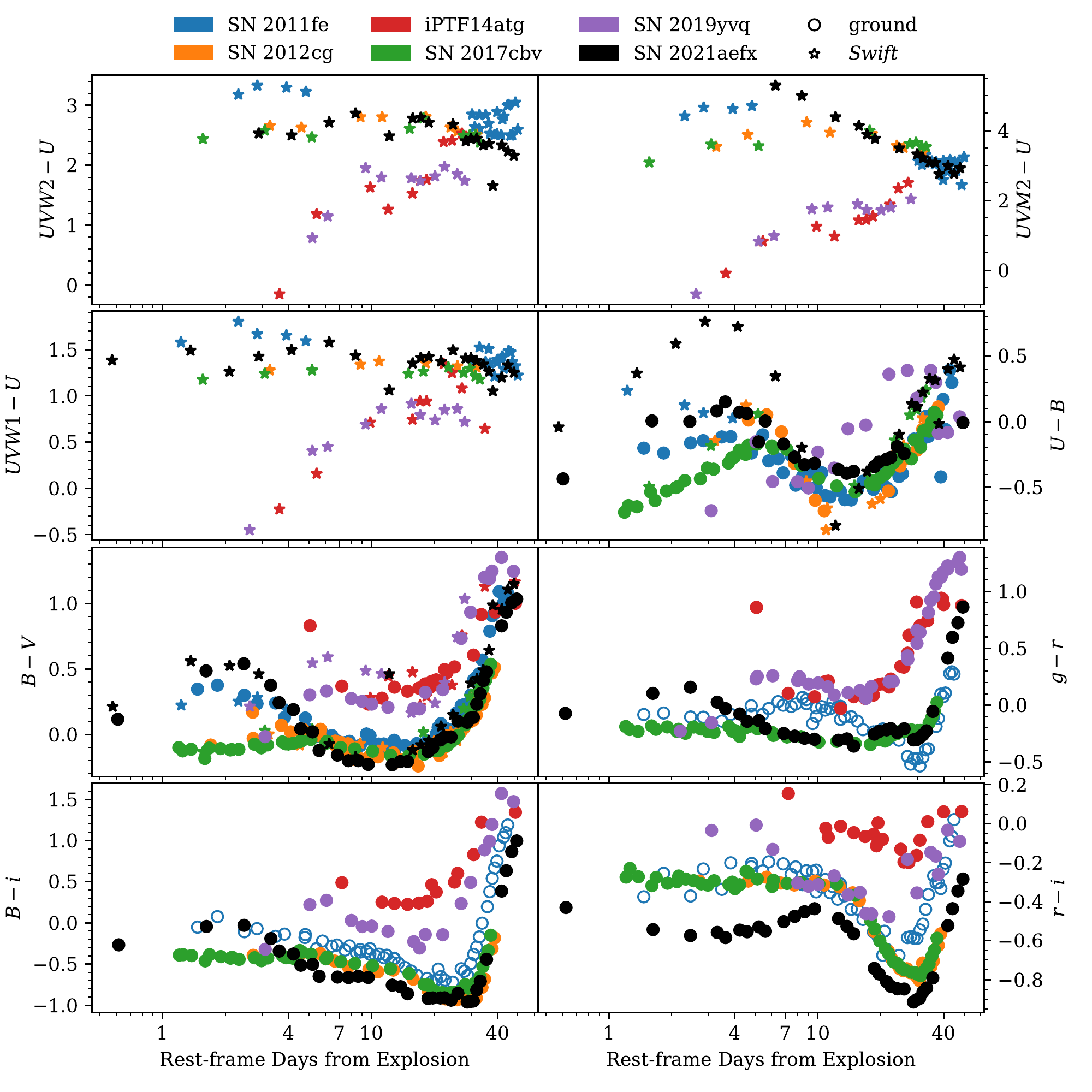}
    \caption{Extinction-corrected color curves of SN~2021aefx compared to SN~2011fe \citep{zhang_optical_2016} and other SNe~Ia with early excesses: SN~2012cg \citep{marion_sn_2016}, iPTF14atg \citep{cao_sn2002es-like_2016}, SN~2017cbv \citep{hosseinzadeh_early_2017}, and SN~2019yvq \citep{burke_bright_2021}. For the purposes of the logarithmic time axis, we adopt explosion estimates from the aforementioned authors and use $\mathrm{MJD}_0 = 59529.19$ (\S\ref{sec:fitting}) for SN~2021aefx. The early color behavior of SN~2021aefx is similar to SN~2017cbv, but the abrupt change in slope (in $U-B$ and other colors) happens earlier (at ${\sim}2$ days instead of ${\sim}5$ days). In the companion-shocking interpretation, this implies a tighter binary and thus a more compact companion \citep{kasen_seeing_2010}.}
    \label{fig:colors}
\end{figure*}

The most striking feature of our early photometry is the ${\sim}2$~day bump in the extremely high-cadence DLT40 unfiltered light curve ($\mathrm{FWHM} = 357{-}871$~nm; Fig.~\ref{fig:lc}, left). As the bump subsides, we also observe a decline in the bluest detected bands ($U$ and \textit{UVW1}; Fig.~\ref{fig:lc}, center) and a rapid reddening in the $U-B$, $B-V$, and $g-r$ color curves (Fig.~\ref{fig:colors}).
Compared to SNe~2017cbv \citep{hosseinzadeh_early_2017} and 2018oh \citep{dimitriadis_k2_2019,li_photometric_2019,shappee_seeing_2019}, which also had well-sampled early bumps, the bump in SN~2021aefx lasts about half as long (2 days vs.\ 5 days). In the following sections, we fit three classes of models to our photometry to explore the cause of the early bump.

Apart from the bump, the light curve of SN~2021aefx is quite typical for an SN~Ia. We fit a fourth-order polynomial to the $B$-band light curve of SN~2021aefx around maximum light (${\sim}5$ days before to ${\sim}20$ days after), using only data from Las Cumbres Observatory's 1\,m telescopes for consistency. The polynomial gives a peak of $M_B = -19.63 \pm 0.02$~mag at MJD 59546.54. We adopt this as phase = 0 throughout our figures and tables. The light curve declines by $\Delta m_{15}(B) = 0.90 \pm 0.02$~mag over the 15 days after that peak. This is consistent with the slow and luminous end of the \cite{phillips_absolute_1993} relation, which gives confidence in the distance we adopt above, despite its large uncertainty. We note in passing that the SALT2 \citep{guy_salt2:_2007} and MLCS2k2 \citep{jha_improved_2007} SN~Ia templates also provide good fits to the optical (\textit{BgVri}) light curves of SN~2021aefx around maximum light.

\subsection{Companion-shocking Model\label{sec:fitting}}
\begin{figure*}
    \centering
    \includegraphics[width=0.49\textwidth]{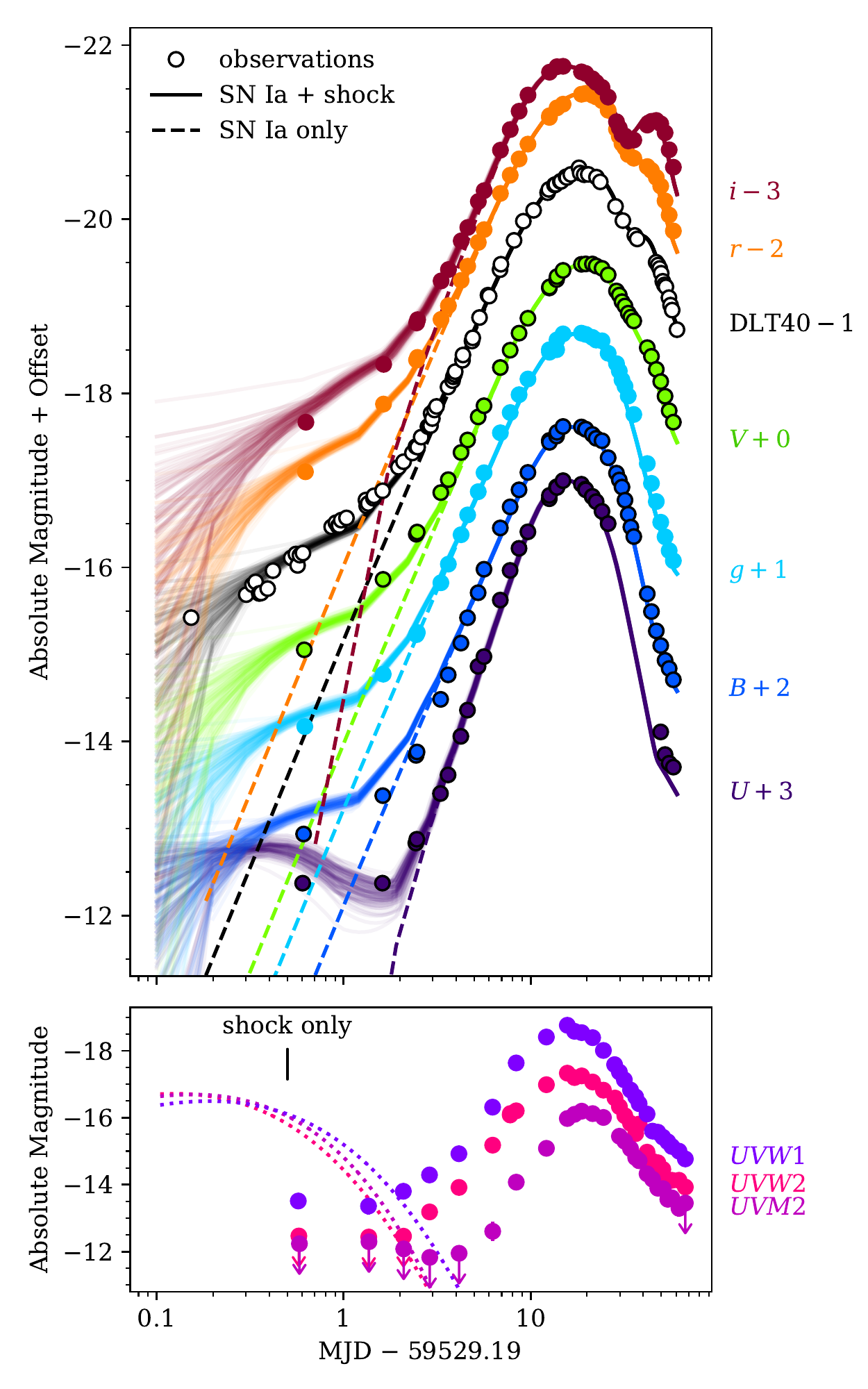}
    \includegraphics[width=0.49\textwidth]{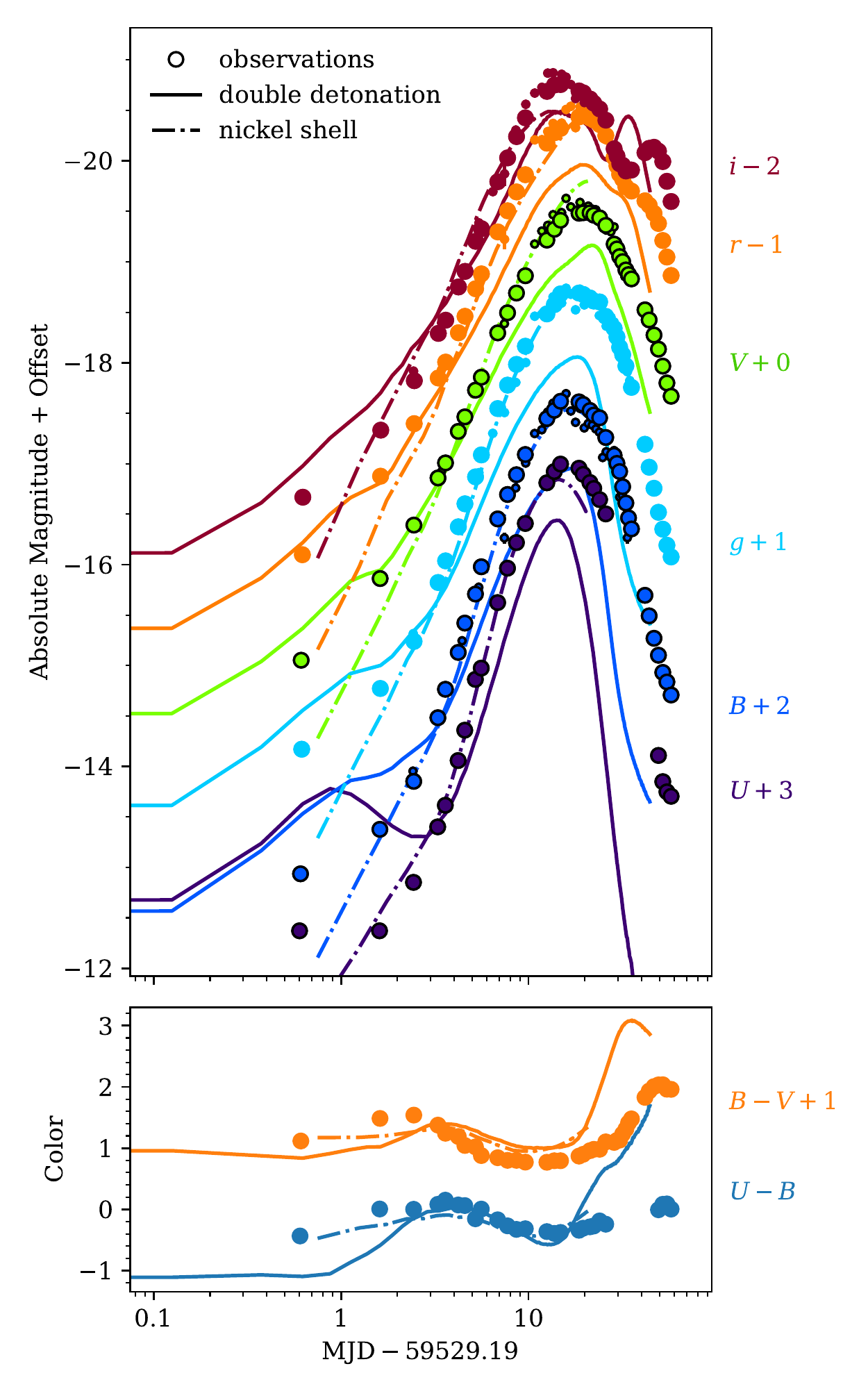}
    \caption{Left: fitting the companion-shocking model of \cite{kasen_seeing_2010} to the light curve of SN~2021aefx. The solid lines in the top panel show the best-fit SiFTO+Kasen model, with the SiFTO model alone shown with dashed lines. The model reproduces the early optical light curve quite well, but the shock component alone (SiFTO does not extend into the UV) significantly overestimates the early UV flux (bottom).
    Right: best-fit edge-lit double-detonation model from \citet[][solid lines]{polin_observational_2019} and nickel-shell model from \citet[][dotted-dashed lines]{magee_investigation_2020} compared to the observed light curve (top) and $U-B$ and $B-V$ color curves (bottom) of SN~2021aefx. This double-detonation model is the explosion of a $0.9 \msun$ WD with a $0.08 \msun$ helium shell. The model reproduces the early color well but overestimates the absolute magnitude of the bump, while also failing to reproduce the peak absolute magnitude and postpeak colors. The nickel-shell model is a fiducial model of SN~2017cbv plus a shell of $0.02 \msun$ of $^{56}$Ni centered on a mass coordinate of $1.35 \msun$ with a width of $0.18 \msun$. This model reproduces the peak better but does not show a strong enough bump in the light curves or color curves. \explain{Shifted double detonation model by half a day.}}
    \label{fig:fitting}
\end{figure*}

\cite{kasen_seeing_2010} predicts that this type of early light-curve bump can arise when the SN ejecta collide with a nondegenerate binary companion, which shock-heats the ejecta and causes them to briefly become bluer and more luminous.
We fit the light curve of SN~2021aefx with a model composed of a SiFTO SN~Ia template \citep{conley_sifto:_2008} and a companion-shocking component from \cite{kasen_seeing_2010}, following the procedure of \cite{hosseinzadeh_early_2017} now released as part of the Light Curve Fitting package \citep{hosseinzadeh_light_2020}.
While the effect of companion shocking depends on the viewing angle, here we employ the formulae for the ``isotropic equivalent luminosity,'' which is approximately the strength of the effect for viewing angles less than 30\degr--40\degr\ from the binary axis \citep{kasen_seeing_2010}.
In performing this fit, we exclude data from Las Cumbres Observatory's 0.4\,m telescopes, which have greater scatter, and we assume that the unfiltered DLT40 light curve can be approximately fit by the $r$-band SiFTO template.
(We confirmed that the early DLT40 and $r$-band light curves are in good agreement.)
Unlike \cite{hosseinzadeh_early_2017}, we do not require additional multiplicative factors on the $r$ and $i$ SiFTO templates, nor on the $U$ shock component, to achieve a reasonable fit.
Instead, we fit for small time shifts in the $U$ and $i$ SiFTO templates to improve the match during the light-curve rise (after the bump).
We also include an intrinsic scatter term $\sigma$, such that the effective uncertainty on each point is increased by a factor of $\sqrt{1+\sigma^2}$.
This allows for scatter not accounted for by the model, as well as photometric uncertainties that are potentially underestimated, in order to estimate more realistic uncertainties on the model parameters.
The model parameters and their respective priors and best-fit values are listed in Table~\ref{tab:priors}.
The light curve and best-fit model are displayed in Fig.~\ref{fig:fitting} (left).

\begin{deluxetable*}{lLlllLl}
\tablecaption{Model Parameters\label{tab:priors}}
\tablehead{\colhead{Parameter} & \colhead{Variable} & \colhead{Prior Shape} & \twocolhead{Prior Parameters\tablenotemark{a}} & \colhead{Best-fit Value\tablenotemark{b}} & \colhead{Units}}
\tablecolumns{7}
\startdata
\cutinhead{Companion-shocking Model \citep{kasen_seeing_2010}}
Explosion time & t_0 & Uniform & 59526.34 & 59529.34 & 59529.19^{+0.07}_{-0.12} & MJD \\
Binary separation & a & Uniform & 0 & 1 & 0.012^{+0.004}_{-0.003} & $10^{13}$ cm \\
Ejecta mass $\times$ velocity$^7$ & Mv^7 & Log-uniform & 0.1 & 500 & 90^{+120}_{-50} & $M_\mathrm{Ch}$ ($10^4$ km s$^{-1}$)$^7$ \\
\cutinhead{SiFTO Model \citep{conley_sifto:_2008}}
Time of $B$ maximum & t_\mathrm{max} & Uniform & 59542.09 & 59548.09 & 59546.54 \pm 0.03 & MJD \\
Stretch & s & Log-uniform & 0.5 & 2 & 1.010 \pm 0.002 & dimensionless \\
Time shift in $U$ & \Delta t_U & Gaussian & 0 & 1 & +0.71^{+0.08}_{-0.07} & d \\
Time shift in $i$ & \Delta t_i & Gaussian & 0 & 1 & +0.52 \pm 0.10 & d \\
\cutinhead{Combined Model}
Intrinsic scatter & \sigma & Half-Gaussian & 0 & 1 & 12.9 \pm 0.4 & dimensionless
\enddata
\tablenotetext{a}{The ``Prior Parameters'' column lists the minimum and maximum for a uniform distribution, and the mean and standard deviation for a Gaussian distribution.}
\tablenotetext{b}{The ``Best-fit Value'' column is determined from the 16th, 50th, and 84th percentiles of the posterior distribution, i.e., $\mathrm{median} \pm 1\sigma$.}
\end{deluxetable*}\vspace{-24pt}

The best-fit binary separation is $a = 1.8\ R_\sun$. If we assume the companion is in Roche lobe overflow \citep[$\frac{a}{R} \approx 2{-}3$;][]{eggleton_approximations_1983,kasen_seeing_2010}, this implies a companion radius of $R \approx 0.7\ R_\sun$.
This and the other values are degenerate with the viewing angle, which we do not account for here (see above), but nonetheless, the order of magnitude suggests a low-mass main-sequence star.
The best-fit explosion time ($t_0 = 59529.19^{+0.07}_{-0.12}$) indicates that our discovery image was taken 3.5 rest-frame hours after explosion.
The remaining parameters are ``nuisance'' parameters for our analysis, but we note that the artificial time shifts we applied to certain bands of the SiFTO model are small (${<}1$~day) and that the stretch is close to 1, confirming again that this is a typical SN~Ia.

As in SN~2017cbv \citep{hosseinzadeh_early_2017}, which showed a similar early bump, the best-fit model drastically overpredicts the UV luminosity (Fig.~\ref{fig:fitting}, bottom right).
The shock component alone is 10, ${>}28$, and ${>}18$ times stronger than the observed luminosity in the \textit{UVW1}, \textit{UVM2}, and \textit{UVW2} bands, respectively.
However, unlike in SN~2017cbv, our high-cadence early Swift light curve shows an initial decline in \textit{UVW1}.
In other words, there is a significant UV bump, but it is not as strong as the model predicts.

\subsection{Sub-Chandrasekhar-mass Models\label{sec:dd}}
\cite{polin_observational_2019} also predict similar early bumps for explosions of sub-Chandrasekhar-mass WDs arising from the double-detonation mechanism.
For each of the 39 models in their grid, we apply time shifts between 0 and 3 days, in half-day increments, between their explosion time and our explosion time estimate from \S\ref{sec:fitting}. We then calculate the reduced $\chi^2$ ($\bar\chi^2$) statistic for each shifted model compared to our \textit{UBgVri} light curve from the Las Cumbres Observatory 1\,m telescopes. Several models have a reasonable fit around peak, the best being a 1.1\msun WD with a 0.01\msun helium shell ($\bar\chi^2=1083$; number of observations $n=52$), but this model has no early bump at all, underpredicting our first epoch of observations by several magnitudes.

To search for models with a good fit to the bump specifically, we repeat this procedure with the $U-B$ color curve. The best-fit model in this case ($\bar\chi^2 = 579$; $n=22$) is shown in Fig.~\ref{fig:fitting} (right) with a time shift of 0.5 days. This model is an edge-lit double detonation of a $0.9 \msun$ WD with a $0.08 \msun$ helium shell. The morphology of the bump is a good fit, with very blue but rapidly reddening colors for ${\sim}2$ days. However, as in the companion-shocking model, this sub-Chandrasekhar-mass model severely overpredicts the bump in the $U$ band. At the same time, it does not reach the relatively bright peak magnitude that we observe in any band, and the colors are much redder than observed after maximum light.

\subsection{Nickel-shell Models\label{sec:ni}}
We also compared the light curves and color curves of SN~2021aefx to the models of \cite{magee_determining_2020}, in which they explode WDs with radioactive $^{56}$Ni mixed into the outer layers using the TURTLS radiative transfer code \citep{magee_modelling_2018}. We repeat the time shift and $\bar\chi^2$ calculation as in \S\ref{sec:dd}. Again we are able to find good fits around peak. However, none of these provide a good fit to the early bump. In many cases, this is because the models only begin ${>}2$ days after explosion. However, as noted by \cite{magee_determining_2020}, most models in their grid do not show an early bump at all.

A more promising set of nickel models come from \cite{magee_investigation_2020}, who set out to reproduce the early bumps in SNe~2017cbv and 2018oh. These models include a shell of nickel artificially placed into the outer ejecta. The model set is smaller and tailored to two specific SNe, but nonetheless we repeat the time shift and $\bar\chi^2$ calculation, this time using both $U-B$ and $B-V$ colors. We find that the SN~2017cbv model with a shell of $0.02 \msun$ of $^{56}$Ni centered on a mass coordinate of $1.35 \msun$ with a width of $0.18 \msun$ provides a reasonable fit to both the light curves ($\bar\chi^2 = 205$; $n=187$) and color curves ($\bar\chi^2 = 211$; $n=34$) of SN~2021aefx, with no time shift. Although this model provides a better fit to the light-curve peak than the double-detonation model, the early bump is not strong enough to match our first epoch of observations. Further modeling tailored to SN~2021aefx could potentially provide a better fit. However, \cite{magee_investigation_2020} also note that their models greatly overpredict the strength of the UV bump. We discuss all the above models further in \S\ref{sec:discuss}.

\subsection{Circumstellar Interaction}
\cite{piro_exploring_2016} explore the possibility that early excesses in SNe~Ia arise from nickel mixing into the outer layers plus a small amount of circumstellar material (CSM). We repeated our fitting process for these models and found some reasonable fits to the $U-B$ and $B-V$ color curves: all the models with 0.3\msun of CSM and mixing widths of 0.05\msun \citep[Fig.~13 of][]{piro_exploring_2016} have $\bar\chi^2 \approx 500$ ($n=22$). However, these models are very poor fits to the light curve; the bumps are much more extreme than we observe. In general, the bumps in these models are not blue enough, in the sense that it is not possible to produce a strong enough bump in the $U$ band without producing almost equal-strength bumps in the redder bands. (Note that their figures show $V$-band bumps, whereas the bump in SN~2021aefx is almost invisible in $V$.)

\subsection{Doppler Shift}
\cite{ashall_speed_2022} recently proposed that some UV bumps can be explained simply by shifting common SN~Ia spectral features into and out of observed filter passbands. In particular, they claim that the $u$-band bump in SN~2021aefx is caused by the \ion{Ca}{2} H\&K absorption feature shifting to wavelengths lower than the peak sensitivity of the $u$ filter, due to the very high expansion velocities in its earliest spectrum (a preliminary reduction of the same data we present here). However, this mechanism can only explain bumps in filters where there is a strong spectral slope across the passband. \cite{ashall_speed_2022} suggest that \textit{UVM2} is another such band, due to the strong absorption from iron-peak elements around 200--250~nm, but they do not discuss the Swift photometry of SN~2021aefx specifically, which shows the bump in \textit{UVW1}.

Importantly, this mechanism cannot explain the bump in our early unfiltered light curve. Whereas shifting the \ion{Ca}{2} H\&K feature to shorter wavelengths can significantly increase the $u$/$U$-band flux, it barely affects the unfiltered optical light curve, even though the CCD has some sensitivity in the UV, as it is averaged over a much broader wavelength range. As a demonstration, we follow the procedure of \cite{ashall_speed_2022} to calculate synthetic photometry\footnote{Rather than extrapolate the spectrum into the UV and infrared to cover the DLT40 bandpass, we only integrate over wavelengths covered by our spectrum. This is equivalent to assuming that the transmission-weighted average $F_\nu$ is the same outside our wavelength coverage as it is inside.} from our early spectrum both as observed and redshifted\footnote{We do this to compare with their results, although we agree with their caveat that simply redshifting a spectrum is not the correct way to simulate lower-velocity lines.} by 16,000~km~s$^{-1}$. We find that the spectrum as observed is 0.55~mag brighter in $U$ and ${<}0.01$~mag brighter in the DLT40 band than the lower-velocity version. In other words, the bump in the unfiltered light curve must also include contributions from the redder bands.

As an additional precaution, we repeated our fit of the companion-shocking model from \S\ref{sec:fitting} without the $U$-band light curve. (The Swift photometry was already not included.) Our best-fit model is nearly identical to the one shown in Fig.~\ref{fig:fitting} (left), including the value of the binary separation. This confirms that, even if the $U$-band photometry is ``contaminated'' by the \ion{Ca}{2} H\&K Doppler effect, the optical light curves alone lead to the same conclusions.

\begin{figure}
    \centering
    \includegraphics[width=\columnwidth]{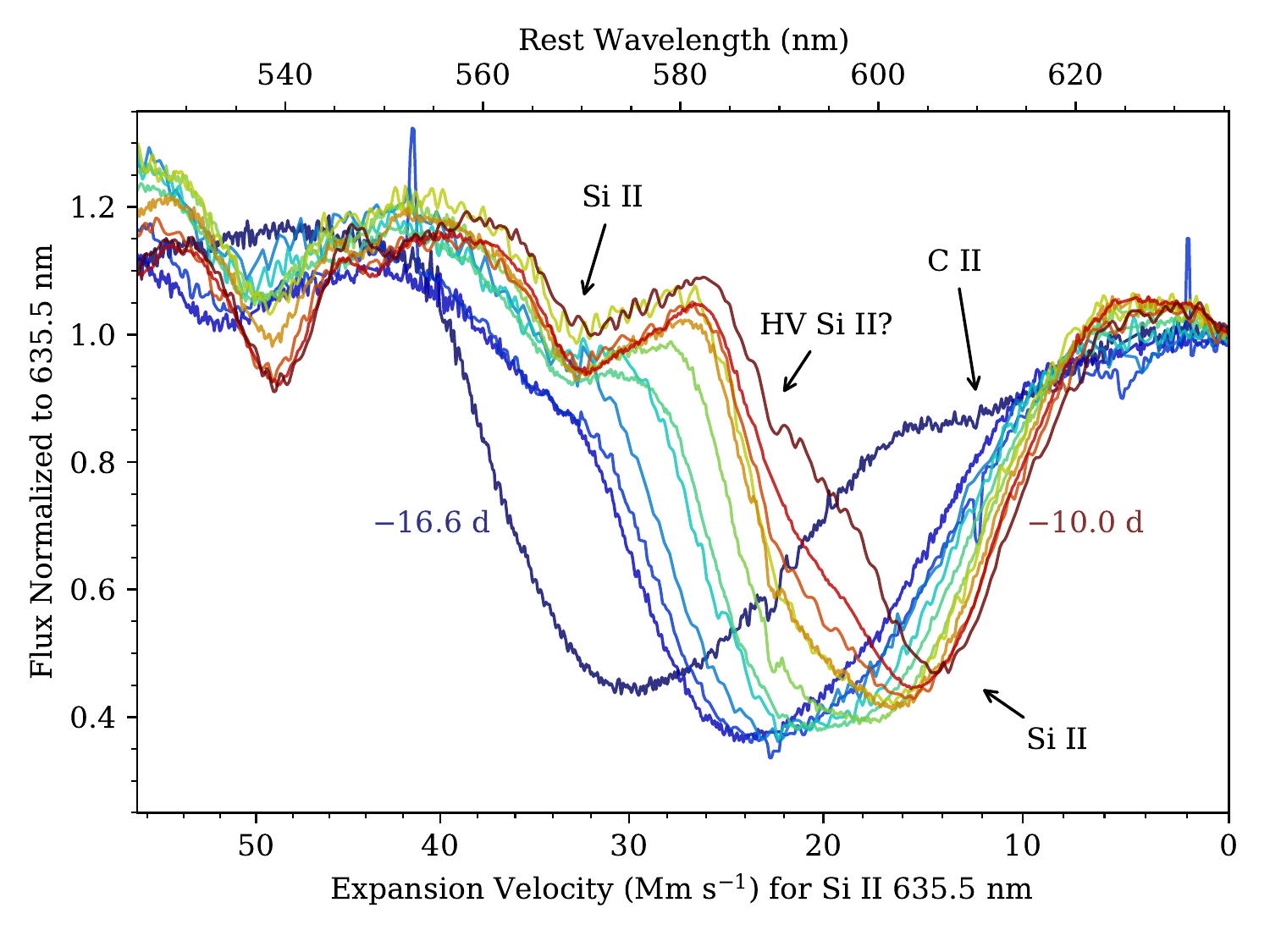}
    \caption{Line identification in the premaximum spectra of SN~2021aefx. Colors are a nonlinear function of time; exact phases are listed in Table~\ref{tab:spec}. Our initial spectrum ($-16.6$ days, dark purple) is dominated by a very broad absorption feature around 570~nm (in the rest frame of the host galaxy). By comparison to later spectra (e.g., at $-10$ days, dark red), this line likely consists of \ion{Si}{2} 597.2~nm, \ion{Si}{2} 635.5~nm, and \ion{C}{2} 658.0~nm. There may also be a high-velocity (HV) component of \ion{Si}{2} 635.5~nm that makes the absorption profile flatter than a Gaussian. The presence of unburned carbon in the early spectrum disfavors the sub-Chandrasekhar-mass double-detonation model for SN~2021aefx.}
    \label{fig:spec_early}
\end{figure}

\section{Early Spectra\label{sec:spec_early}}
The initial spectrum of SN~2021aefx shows an unusually deep and broad absorption feature around 570~nm in the rest frame of the host galaxy. We show a detail of this feature compared to our other premaximum spectra in Fig.~\ref{fig:spec_early}. If we attribute this line to \ion{Si}{2} at 635.5~nm, we infer a very high expansion velocity of ${\sim}29,500$~km~s$^{-1}$. However, the non-Gaussian shape of the line, even at later phases, may indicate the presence of two velocity components, as well as blending with another \ion{Si}{2} line at 597.2~nm. We also observe a second absorption component on the red side of this feature, which we attribute to \ion{C}{2} 658.0~nm. Only ${\sim}10{-}30\%$ of SNe~Ia show carbon in their early optical spectra \citep{parrent_study_2011,folatelli_unburned_2012,silverman_berkeley_2012a,wyatt_strong_2021}, and this feature in SN~2021aefx is one of the strongest observed to date. Unburned carbon has been interpreted as a signature of asymmetry or clumpiness in the WD explosion \citep{hoflich_thermonuclear_2002,ropke_off-center_2007,shen_ignition_2014}, which serves as a probe of the explosion mechanism. For example, unburned carbon is not expected for Chandrasekhar-mass delayed-detonation models \citep{kasen_diversity_2009} or from sub-Chandrasekhar-mass double-detonation models \citep{fink_double-detonation_2010,polin_observational_2019}. Several other SNe~Ia with early light-curve excesses have exhibited strong, early carbon signatures, including iPTF16abc \citep{miller_early_2018}, SN~2017cbv \citep{hosseinzadeh_early_2017}, and SN~2018oh \citep{li_photometric_2019}, a suggestive feature that we will explore in future work.

Our spectrum near maximum light ($+4$ days) closely resembles near-maximum spectra of several normal SNe~Ia, particularly SN~1981B, which spectroscopic classifiers Gelato \citep{harutyunyan_esc_2008}, SNID \citep{blondin_determining_2007}, and Superfit \citep{howell_gemini_2005} all place in the top three best matches. In this spectrum, where the lines are not blended, we fit a linear continuum minus a Gaussian to the \ion{Si}{2} lines at 635.5~nm and 597.2~nm. We measure equivalent widths of $10.6 \pm 0.7$~nm and $1.7 \pm 0.6$~nm, respectively, both at a velocity of $11\,390 \pm 30$~km~s$^{-1}$. This places it near the boundary of the ``core normal'' and ``broad line'' subclasses of \cite{branch_comparative_2006} and in the high-velocity subclass of \cite{wang_evidence_2013}. Along with the light-curve widths in \S\ref{sec:phot} ($\Delta m_{15}(B) = 0.90$~mag) and \S\ref{sec:fitting} ($s = 1.010$), these measurements indicate that SN~2021aefx is a typical SN~Ia that could be used for cosmology. More detailed modeling of the unique early spectrum, along with a full spectroscopic analysis, will be presented by L.~A.~Kwok et al.\ (2022, in preparation).

\section{Nebular Spectroscopy\label{sec:nebular}}
We measure complementary constraints on the progenitor system of SN~2021aefx using the late-time SOAR spectrum (117.9~days after $B$-band maximum), which was taken with the Goodman Spectrograph's red camera, 600 line mm$^{-1}$ grating, and a $1''$ slit, resulting in an $R \approx 1250$.  If the progenitor system of SN~2021aefx had a nondegenerate companion star, as the early light curve hints at, then models predict that the SN ejecta will ablate the companion and lead to narrow hydrogen or helium emission lines ($\mathrm{FWHM} \approx 1000$ km s$^{-1}$) $\gtrsim$100 days after the explosion \citep[e.g. most recently][]{botyanszki_multidimensional_2018,dessart_spectral_2020}. The models predict ${\gtrsim}0.1 M_{\odot}$ of stripped hydrogen, although there is diversity in the expected strength and shape of the resulting emission line, which will depend on the details of the explosion, nondegenerate companion, and radiation transfer physics employed.

Direct inspection of the SOAR spectrum reveals no hydrogen or helium emission features, and we set quantitative limits on any narrow H$\alpha$ or \ion{He}{1} (either $\lambda$5875~\AA\ and $\lambda$6678~\AA) emission using the methodology of \citet{sand_nebular_2018,sand_nebular_2019}.  Briefly, we take the flux-calibrated  and extinction-corrected spectrum and bin to the native resolution.  Following this, we smoothed the spectrum on scales larger than the expected emission ($\mathrm{FWHM} \approx 1000$ km s$^{-1}$) using a second-order \cite{savitzky_smoothing_1964} filter with a width of 130~\AA.  Any hydrogen or helium feature with a width similar to that expected in the single degenerate scenario would be apparent in the difference between the unsmoothed and smoothed versions of the spectrum.

To set quantitative limits on our H$\alpha$ and \ion{He}{1} nondetections, we directly implant emission lines into our data, assuming a line width of 1000 km s$^{-1}$ and a peak flux that is three times the root mean square of the residual spectrum described above.  This results in flux and luminosity (assuming $D=18.0$ Mpc) limits for i) H$\alpha$ of 1.6$\times$10$^{-15}$ erg s$^{-1}$ cm$^{-2}$ and 6.3$\times$10$^{37}$ erg s$^{-1}$; ii) \ion{He}{1} $\lambda$5875~\AA\ of 4.8$\times$10$^{-15}$ erg s$^{-1}$ cm$^{-2}$ and 1.9$\times$10$^{38}$ erg s$^{-1}$; and iii) \ion{He}{1} $\lambda$6678~\AA\ of 1.6$\times$10$^{-15}$ erg s$^{-1}$ cm$^{-2}$ and 6.3$\times$10$^{37}$ erg s$^{-1}$, respectively.  These limits are also listed in Table~\ref{tab:results}.  This H$\alpha$ luminosity limit is comparable to, or fainter than, the recent detections identified in three fast-declining SNe Ia \citep{kollmeier_h_2019,prieto_variable_2020,elias-rosa_nebular_2021}.

To translate these hydrogen and helium luminosity limits to masses, we utilize the radiation transport model sets of \citet{botyanszki_multidimensional_2018} and \citet{dessart_spectral_2020} and refer the reader directly to those works for details.  In brief, the 3D radiation transport results of \citet{botyanszki_multidimensional_2018} presented simulated SN Ia spectra at 200~days after explosion, based on the SN~Ia ejecta--companion interaction simulations of \citet{boehner_imprints_2017}.  The simulated spectra show $L_\mathrm{H\alpha} \approx (4.5{-}15.7) \times 10^{39}$ erg s$^{-1}$ for their main-sequence, subgiant, and red-giant companion star models (corresponding to $M_\mathrm{strip} \sim 0.2{-}0.4 \msun$), roughly two orders of magnitude brighter than our detection limit.  To quantify our stripped mass limit, we  adopt the quadratic fitting relation between H$\alpha$ (and \ion{He}{1} $\lambda$5875~\AA\ and $\lambda$6678~\AA) luminosity and stripped hydrogen/helium mass (Eq.~1 in \citealt{botyanszki_multidimensional_2018}, updated by \citealt{sand_nebular_2018}), and make a correction for the epoch of our observations (117.9~days after $B$-band max, or 135.2~days after explosion), similar to that done in \citet{botyanszki_multidimensional_2018}.  This leads to a stripped hydrogen mass limit of $M_\mathrm{H,strip} \lesssim 4.5 \times 10^{-4} \msun$, and a helium mass limit of $M_\mathrm{He,strip} \lesssim 5.4 \times 10^{-3} \msun$ and $M_\mathrm{He,strip} \lesssim 2.4 \times 10^{-3} \msun$ for the \ion{He}{1} $\lambda$5875~\AA\ and $\lambda$6678~\AA\ lines, respectively.

We also constrain the amount of stripped hydrogen based on 1D radiative transfer calculations from several different delayed-detonation models \citep[DDC0, DDC15, and DDC25;][]{blondin_one-dimensional_2013}, including non-LTE physics and optical depth effects, as presented by \citet{dessart_spectral_2020}.  Here we take the available model grids, which run from 100 to 300~days after explosion and interpolate them to the explosion epoch of SN~2021aefx.  With these models, our stripped hydrogen mass limit is between $M_\mathrm{H,strip} \lesssim (5{-}6) \times 10^{-4} \msun$, depending on the model used.  These same radiative transfer simulations produce weak or ambiguous helium emission lines (aside from the \ion{He}{1} $\lambda$10830~\AA\ line, which is unavailable to us), so we do not calculate stripped helium mass limits for this model framework. We tabulate our hydrogen and helium luminosity and stripped mass limits in Table~\ref{tab:results}.

We recommend the collection of a sequence of nebular spectra to further constrain and search for signatures of the single-degenerate scenario out to later times.  Other works have suggested searching for other narrow emission lines, such as [\ion{O}{1}]$\lambda$6300~\AA\ and \ion{Ca}{2}$\lambda$$\lambda$7291,7324~\AA, which may be sensitive to helium star companions in particular \citep[e.g.,][]{lundqvist_hydrogen_2013}.  While our deep SOAR spectrum constrains the luminosity of the [\ion{O}{1}]$\lambda$6300~\AA\ line to be $L_\mathrm{[O~I]} < 6.3 \times 10^{37}$ erg s$^{-1}$, data at later times will further constrain the progenitor system of SN~2021aefx.

\section{Radio Nondetection\label{sec:radio}}
We model the radio emission from the SN in the same way as \citet{lundqvist_deepest_2020}. That is, the emission arises as a
result of circumstellar interaction when electrons  are accelerated to relativistic speeds behind the supernova blast wave where
significant magnetic fields are also generated. The relativistic electrons radiate synchrotron emission. For a power-law distribution
of the electron energies, $dN/dE = N_0E^{-p}$, where $E=\gamma m_ec^2$ is the energy of the electrons and $\gamma$ is the Lorentz factor,
the intensity of optically thin synchrotron emission $\propto \nu^{-\alpha}$, where $\alpha = (p-1)/2$. We have used $p=3$, as this has been
shown to be a good choice for SNe~Ibc \citep{chevalier_circumstellar_2006}. 
 
The structure of the CSM depends on the progenitor system. For a double-degenerate progenitor system, one may expect 
a constant-density medium (see below), whereas for most single-degenerate systems, some sort of wind mass loss is likely to 
occur. We probe the latter by assuming a constant mass-loss rate, $\dot M$, and a constant wind speed, $v_w$. For this scenario, the density of 
the CSM decreases with distance from the progenitor system as $\rho(r) = \dot M/(4\pi r^2 v_w)$. 

For the SN ejecta, we employ a model called N100 \citep{ropke_constraining_2012,seitenzahl_three-dimensional_2013} to test the single-degenerate scenario, as was also used by
\citet{lundqvist_deepest_2020}. This is a delayed-detonation model where the ignition occurs in 100 sparks in the central region. The total mass and 
asymptotic kinetic energy of the ejecta for N100 are $1.4 \msun$ and $1.45\EE{51}$ erg, respectively. The density profile of the ejecta in N100
is only given for velocities ${\leq}2.8 \times 10^{4}\ \kms$. For faster ejecta, a power-law density structure, $\rho_\mathrm{ej} \propto r^{-n}$,
was added. As in the models of \citet{lundqvist_deepest_2020}, who modeled radio upper limits for several nearby SNe~Ia, we have assumed 
$n = 13$ \citep[see][for a discussion on $n$]{kundu_constraining_2017}. The same value for $n$ facilitates comparison with those nearby supernovae.
For the same reason, we have also assumed the same value for the brightness temperature of the radio emission at the frequency 
at which the optical depth of synchrotron self-absorption is unity as in \citet{lundqvist_deepest_2020}, namely $T_{\rm bright} = 5\EE{10}~{\rm K}$ 
(see \citealt{bjornsson_heating_2014} for further discussion on $T_{\rm bright}$). 
\begin{figure}
\centering 
\includegraphics[width=\columnwidth]{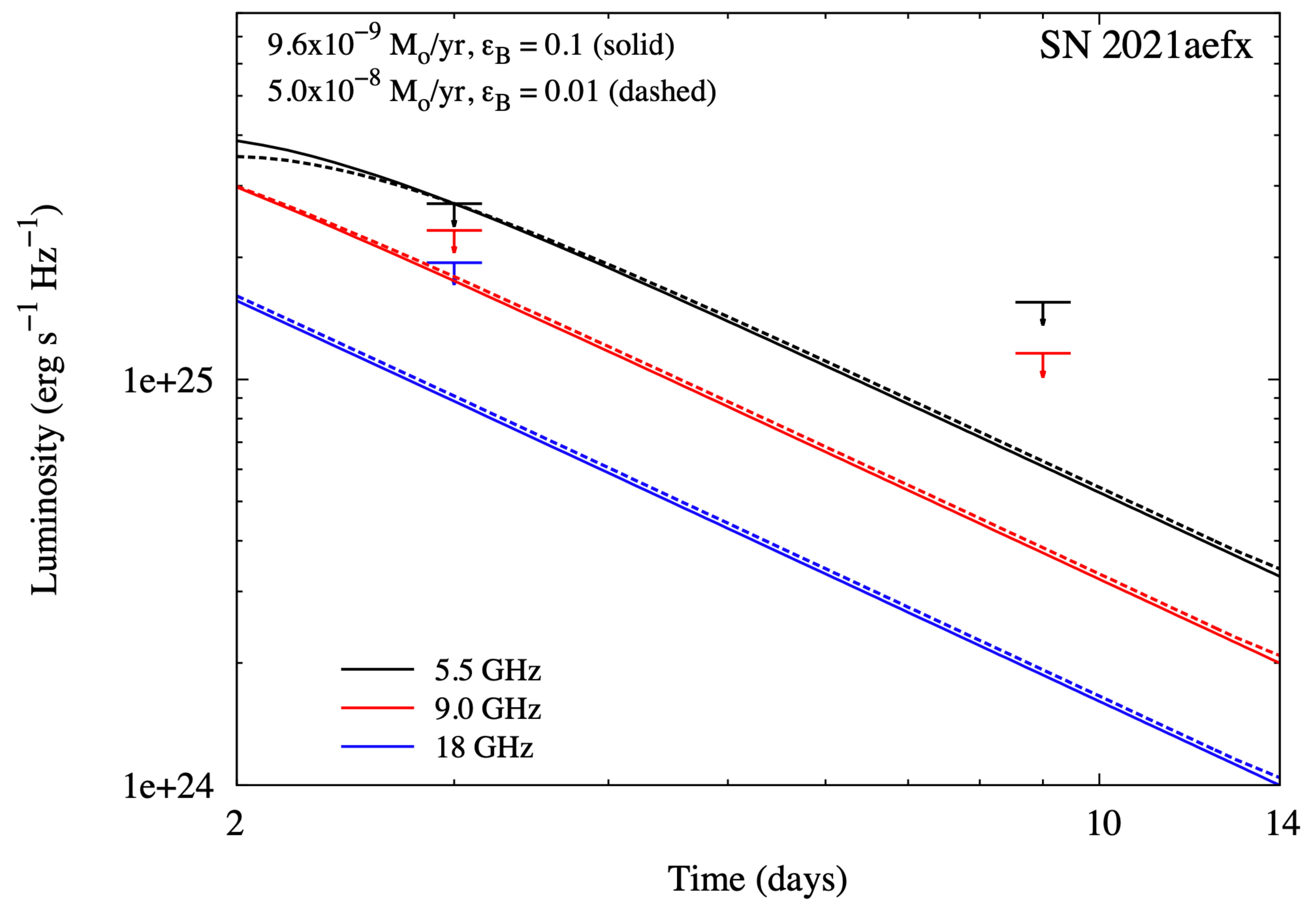}
\caption{
Radio data for SN~2021aefx \citep{kundu_mass-loss_2021} compared to
models at various frequencies for a $\rho(r) \propto r^{-2}$ wind. Models use 
$\dot{M} = 0.96~(5.0) \EE{-8} \msunyr~(v_w/100 \kms)$
for $\epsilon_B = 0.1~(0.01)$, with solid lines being for
$\epsilon_B = 0.1$. Common parameters in both models
are $\epsilon_{\rm e} = 0.1$, $T_{\rm bright} = 5\EE{10}~{\rm K}$,
$n=13$ and $v_w = 100~\kms$. The most constraining observation is at 5.5 GHz on 
day 3.
}
\label{fig:21aefx_fig}
\end{figure}

In Fig.~\ref{fig:21aefx_fig} we show the predicted radio emission from our model. Using the best-fit explosion time from \S\ref{sec:fitting},
the two epochs of radio observations occurred 3 and 9 days after explosion.
The ratio of $\dot M/v_w$ has been tuned to give the highest luminosity possible that does not conflict with observational limits. Solid lines are 
for high efficacy of conversion of shock energy to magnetic field energy density in the postshock region behind the supernova blast wave. If we define 
$\epsilon_B = u_B / u_{\rm th}$, where $u_{\rm th}$ is the postshock thermal energy density and  $u_B$ is the magnetic field 
energy density, $\epsilon_B = 0.1$ in this model. Dashed lines are for $\epsilon_B = 0.01$. (In both models we have used  
$\epsilon_{\rm rel} = 0.1$, where $\epsilon_{\rm rel} = u_{\rm rel} / u_{\rm th}$, and $u_{\rm rel}$ is the energy density in relativistic electrons.)  
As can be seen, a higher wind density is needed to compensate for less efficient conversion to magnetic field energy density. 
For $\epsilon_B = 0.1~(0.01)$ the wind density is described by $\dot{M} = 0.96~(5.0) \EE{-8} \msunyr~(v_w/100 \kms)$. 
For both choices of $\epsilon_B$, the constraining observation is that at 5.5 GHz 3 days after the explosion.
The light curves only differ significantly at $t \lesssim 3$ days in the 5.5 GHz band. Here, synchrotron self-absorption plays a role in the $\epsilon_B = 0.01$ 
 model.
 
From the compilation of nearby SNe~Ia by \citet{lundqvist_deepest_2020}, we note that only one SN~Ia has been younger at the time of first radio observation, 
namely SN~2011fe \citep[see][]{chomiuk_deep_2016}. Fig.~\ref{fig:21aefx_fig} shows that early observations are essential to constrain any wind material. The 
compilation of \citet{lundqvist_deepest_2020} further shows that SN~2021aefx is one of the most constraining cases in that sense, only clearly surpassed by
SNe~2011fe, 2012cg, and 2014J. Despite this, it is difficult to rule out the majority of formation channels of single-degenerate scenarios, even 
if $\epsilon_B$ were as high as ${\sim} 0.1$ \citep[see][]{lundqvist_deepest_2020}. As discussed by \citet{lundqvist_deepest_2020}, we may expect $\epsilon_B$ in the 
range of 0.01--0.1 for young supernovae, but there is considerable uncertainty \citep{reynolds_efficiencies_2021}.
 
In the companion-shocking model, radio emission is likely to be produced as a result of the interaction between the SN and
the binary companion. However, at the time of the first radio observations, the blast wave had already reached $2.3~(1.8)\EE{15}$ cm for
the $\epsilon_B = 0.1~(0.01)$ models. This means that the companion interaction occurred deep inside the ejecta. For example, the outermost ejecta with power-law slope $n=13$ stretch  between $2.5\times 10^{14}$ 
and $2.3 \times 10^{15}$ cm on day 3. Assuming  that this shell mainly consists of carbon, the electron density at its innermost radius is 
${\sim} 3.9\EE8 \cm3$, if carbon is singly ionized. If we further assume a temperature of $10^4$ K, the free--free optical depth through this shell alone
is ${\sim} 1.1\EE4$ at 5.5 GHz. Free--free opacity through ejecta at velocities ${\lesssim} 2.8\times10^{4}\ \kms$ will add to this optical depth. It is
therefore safe to state that no radio emission could have escaped from the companion interaction region on day 3 unless we could view this through 
a hole in the ejecta.
 
For the sub-Chandrasekhar model in \S\ref{sec:dd}, the time between explosion and first radio observation is closer to 4 days. Even if the N100 model
may not be ideal for this case, Fig.~\ref{fig:21aefx_fig} shows that a higher value for $\dot M/v_w$ can be accommodated if the time between
the explosion and the first radio observation is increased. If we use N100 and 4 days after explosion, 
$\dot{M} = 1.23~(6.4) \EE{-8} \msunyr~(v_w/100 \kms)$ for $\epsilon_B = 0.1~(0.01)$, respectively. 
 
For a situation with a constant-density ambient medium, which may be the case for a double-degenerate scenario, the 
expected radio flux increases with time \citep[e.g.,][]{kundu_constraining_2017,lundqvist_deepest_2020}. To test this, we invoke the violent merger model 
of \cite{pakmor_normal_2012}, simulating the merger of two C/O degenerate stars with masses of $1.1 \msun$ and $0.9 \msun$. The total mass and 
asymptotic kinetic energy of the ejecta for this model are $1.95 \msun$ and $1.7\EE{51}$ erg, respectively. If we use 5.5 GHz from our 
second data set and assume that the data probe interaction between the supernova and the ambient medium 10 days after explosion,
the upper limit on the density of the ambient medium is $200~(1180) \cm3$ for $\epsilon_B = 0.1~(0.01)$.

In summary, the radio data cannot probe any possible direct interaction between the ejecta and a binary companion, but they can be used
to put a limit on the density of the ambient medium outside ${\sim} 2\EE{15}$ cm. In a single-degenerate scenario,
the mass-loss rate of the progenitor system is ${\gtrsim} 1\EE{-8} \msunyr~(v_w/100 \kms)$ for $\epsilon_B = 0.1$. For the double-degenerate
scenario, or the single-degenerate scenario with spun-up/spun-down super-Chandrasekhar-mass WDs \citep{distefano_spin-up/spin-down_2011,justham_single-degenerate_2011}, where mass transfer
no longer occurs at the time of explosion, a near-uniform density of the ambient medium is expected. However, the radio data are not ideal to test
this, because the likely density is of order $1 \cm3$, and our upper limit on the density is $200 \cm3$ for $\epsilon_B = 0.1$. The limits on
mass-loss rates and ambient medium density are a factor of ${\sim} 5$ higher for $\epsilon_B = 0.01$.  

\begin{deluxetable}{ll}
\tablecaption{Summary of Results\label{tab:results}}
\tablehead{\colhead{Parameter} & \colhead{Value}}
\startdata
Last nondetection & MJD 59524.328 \\
First detection & MJD 59529.343 \\
Distance & $18.0 \pm 2.0$ Mpc \\
Distance modulus & $31.28 \pm 0.23$ mag \\
Redshift & 0.005017 \\
Milky Way $E(B-V)$ & 0.0079 mag \\
Host-galaxy $E(B-V)$ & 0.097 mag \\
Peak absolute magnitude ($B$) & $-19.63 \pm 0.02$ mag \\
Peak time ($B$) & MJD 59546.54 \\
$\Delta m_{15}(B)$ & $0.90 \pm 0.02$ mag \\
\ion{Si}{2} velocity ($-17$ d) & ${\approx}29,500$ km s$^{-1}$ \\
\ion{Si}{2} velocity (+4 d) & $11,390 \pm 30$ km s$^{-1}$ \\
\ion{Si}{2} 635.5~nm EW (+4 d) & $10.6 \pm 0.7$ nm \\
\ion{Si}{2} 597.5~nm EW (+4 d) & $1.7 \pm 0.6$ nm \\
H$\alpha$ luminosity & ${<}6.3 \times 10^{37}$ erg s$^{-1}$ \\
\ion{He}{1} $\lambda$5875~\AA\ luminosity & ${<}1.9 \times 10^{38}$ erg s$^{-1}$ \\
\ion{He}{1} $\lambda$6678~\AA\ luminosity & ${<}6.3 \times 10^{37}$ erg s$^{-1}$ \\
Stripped H mass (Boty\'anszki) & ${\lesssim}4.5 \times 10^{-4} \msun$ \\
Stripped H mass (Dessart) & ${\lesssim}(5{-}6) \times 10^{-4} \msun$ \\
Stripped He mass (Boty\'anszki) & ${\lesssim}(2{-}5) \times 10^{-3} \msun$ \\
\enddata
\end{deluxetable}
\vspace{-12pt}

\section{Progenitor Constraints\label{sec:discuss}}
Taken together, our constraints from the optical/UV light curves (\S\ref{sec:phot}), nebular spectra (\S\ref{sec:nebular}), and radio observations (\S\ref{sec:radio}) place strong but conflicting constraints on the nature of the progenitor of SN~2021aefx. The nondetections of stripped material and circumstellar interaction decisively favor the double-degenerate scenario. On the other hand, the most recent numerical models of double-degenerate explosion mechanisms, even those with unusual nickel distributions, do not match our observed light curves.

Intriguingly, there are good matches to the bump among the color curves of sub-Chandrasekhar SNe~Ia modeled by \cite{polin_observational_2019}. However, several pieces of evidence work against this progenitor model for SN~2021aefx. (1) The luminosity and \ion{Si}{2} velocity of SN~2021aefx at maximum light place it in \citeauthor{polin_observational_2019}'s class of Chandrasekhar-mass WD explosions.\footnote{However, \cite{shen_multidimensional_2021} can reproduce these observables with a 1.1\msun WD.} (2) \cite{polin_observational_2019} predict that thick helium shells (like our best-fit model; $0.08 \msun$ of helium) would produce nearly featureless spectra during the bump (see their Fig.~8), whereas our spectrum has very strong, broad absorption features (Fig.~\ref{fig:spec_early}). (3) \cite{polin_observational_2019} predict that the spectra at peak would be heavily line blanketed in the blue, which we do not observe (Fig.~\ref{fig:spec}). (4) Unburned carbon is not expected in double detonations \citep{polin_observational_2019}, yet \ion{C}{2} absorption is clearly detected in our premaximum spectra.\footnote{However, \cite{dessart_constraints_2014} do expect unburned carbon in their ``pulsational-delayed-detonation'' models.} (5) \cite{polin_nebular_2021} predict that sub-Chandrasekhar SNe~Ia will show strong [\ion{Ca}{2}] emission in their nebular spectra, which we do not observe, at least up to 134~days after maximum light (Fig.~\ref{fig:spec}). As such, although the bump color roughly matches our observations, we do not favor the double-detonation mechanism for SN~2021aefx.

The companion-shocking model of \cite{kasen_seeing_2010} appears to show a much better fit to the light curve, except for the fact that it greatly overpredicts the UV luminosity of the bump. This is most likely a weakness of the assumption of a blackbody SED for the shock component. While the shock front itself may radiate as a blackbody, much of this light will be reprocessed by the SN ejecta before reaching the observer, imprinting an SN~Ia spectrum, which shows strong absorption lines in the UV, on the shock component as well. (\citealt{hosseinzadeh_early_2017} applied a suppression factor to the $U$-band shock component in SN~2017cbv for this reason.) Modeling the shock emission from SN ejecta colliding with a binary companion using a more realistic SED is a promising avenue for future theoretical work.

When comparing the light curves around maximum light, it is important to remember the very different provenances of these SN~Ia models: The models of \cite{polin_observational_2019} are physical, whereas the SiFTO model of \cite{conley_sifto:_2008} is purely observational. In addition, our fitting procedure for the latter, which scales each band independently and applies time offsets to the two bands where they are necessary, essentially guarantees a good fit to the luminosity and colors at peak. Therefore, our claim of a good fit is not based on the peak of the light curve, but rather on the fact that it is possible for the companion-shocking model to produce a bump with roughly the right shape and timescale in all the optical bands.

Our limits on stripped hydrogen are ${\sim}2{-}3$ orders of magnitude below expectations from recent radiative transfer models for nebular emission in the single-degenerate scenario \citep{botyanszki_multidimensional_2018,dessart_spectral_2020}, which generically predict stripped hydrogen in the range $M_\mathrm{H,strip} \approx 0.1{-}0.5 \msun$.  There are important caveats to these results, as they depend on the physics and limited parameter space explored by the simulations we compare our observations to.  Future work including SNe~Ia with a variety of explosion strengths, companion separations, and companion types would greatly enhance this probe for progenitor studies.  To our knowledge, SN~2021aefx is the third normal SN~Ia with an early light-curve excess that did not show hydrogen/helium emission-line features in their late-time spectra, after SN~2017cbv \citep{sand_nebular_2018} and SN~2018oh \citep{dimitriadis_nebular_2019,tucker_no_2019}. Nebular spectra of the subluminous SN~2019yvq, which also had an early excess, did not show hydrogen/helium emission either \citep{siebert_strong_2020,burke_bright_2021,tucker_sn_2021}.

Our radio limits rule out ejecta--wind interaction for mass-loss rates of order $10^{-8} \msunyr$ or higher (assuming $v_w \approx 100$~km~s$^{-1}$), which excludes all symbiotic systems \citep[e.g.,][]{seaquist_highly_1993}. This is in agreement with \cite{chomiuk_deep_2016}, who showed that ${\lesssim}10\%$ of SNe~Ia come from symbiotic progenitors. They also rule out the majority of single-degenerate systems with high accretion rates, which are expected to give rise to optically-thick disk winds \citep[e.g.,][]{badenes_are_2007}. Though not able to rule out all possible CSM configurations, our limits are among the strongest pieces of evidence of a clean circumbinary environment around the exploding WD, nearly on par with the deep X-ray nondetections of SNe~2017cbv and 2020nlb \citep{sand_circumstellar_2021}.

\section{Summary and Conclusions}
We have presented high-cadence optical and UV photometry of the normal Type~Ia SN~2021aefx showing a UV excess during the first ${\sim}2$ days of observation. This bump can either be explained as a collision between the ejecta and a main-sequence binary companion or as an intrinsic property of a sub-Chandrasekhar-mass explosion. Both of our best-fit models broadly reproduce the light-curve morphology but with specific weaknesses that must be addressed before reaching a strong conclusion. We also presented nebular spectroscopy of SN~2021aefx showing no evidence of hydrogen- or helium-rich material stripped from a nondegenerate binary companion, as well as radio observations showing no evidence for interaction with circumstellar or circumbinary material. These nondetections strongly favor the double-degenerate scenario, but ultimately we are not able to prefer one progenitor scenario over the other given the existing models.

SN~2021aefx joins the growing sample of SNe~Ia with unexpected photometric evolution during the first few days after explosion and highlights the importance of early discovery and classification and high-cadence multiwavelength follow-up in confronting existing models. Importantly, these types of light curves will not be available from the upcoming Legacy Survey of Space and Time at Vera C.\ Rubin Observatory with its nominal observing strategy \citep{ivezic_lsst:_2019}. Specialized high-cadence surveys like DLT40, as well as robotic follow-up with facilities like Las Cumbres Observatory, will continue to be required to make progress in understanding SN~Ia progenitors and explosions.

\facilities{ADS, ATCA, CTIO:PROMPT, LCOGT (SBIG, Sinistro, FLOYDS), Meckering:PROMPT, NED, NTT (EFOSC2), SALT (HRS, RSS), SOAR (Goodman), Swift (UVOT)}

\defcitealias{astropycollaboration_astropy_2018}{Astropy Collaboration 2018}
\software{AIPS \citep{wells_nrao's_1985}, Astropy \citepalias{astropycollaboration_astropy_2018}, BANZAI \citep{mccully_lcogt/banzai:_2018}, CASA \citep{mcmullin_casa_2007}, \texttt{corner} \citep{foreman-mackey_corner.py:_2016}, \texttt{emcee} \citep{foreman-mackey_emcee_2013}, FLOYDS pipeline \citep{valenti_first_2014}, Gelato \citep{harutyunyan_esc_2008}, Goodman pipeline \citep{torres_goodman_2017}, HEASoft \citep{nasaheasarc_heasoft:_2014}, \texttt{lcogtsnpipe} \citep{valenti_diversity_2016}, Light Curve Fitting \citep{hosseinzadeh_light_2020}, Matplotlib \citep{hunter_matplotlib:_2007}, MIRIAD \citep{sault_retrospective_1995}, NumPy \citep{oliphant_guide_2006}, PyRAF \citep{sciencesoftwarebranchatstsci_pyraf:_2012}, PySALT \citep{crawford_pysalt:_2010}, SALT HRS MIDAS pipeline \citep{kniazev_mn48:_2016,kniazev_salt_2017}, SNCosmo \citep{barbary_sncosmo_2022}, SNID \citep{blondin_determining_2007}, Superfit \citep{howell_gemini_2005}}

\vspace{6pt}
G.H.\ thanks A.~L.~Piro for sharing his SN~Ia models.
D.J.S.\ would like to thank P.\ Brown for help with planning the Swift follow-up observing campaign.
Time-domain research by the University of Arizona team and D.J.S.\ is supported by NSF grants AST-1821987, 1813466, 1908972, and 2108032, and by the Heising-Simons Foundation under grant \#2020-1864. P.L.\ acknowledges support from the Swedish Research Council. J.E.A.\ is supported by the international Gemini Observatory, a program of NSF's NOIRLab, which is managed by the Association of Universities for Research in Astronomy (AURA) under a cooperative agreement with the National Science Foundation, on behalf of the Gemini partnership of Argentina, Brazil, Canada, Chile, the Republic of Korea, and the United States of America. Research by Y.D., N.M., and S.V.\ is supported by NSF grants AST-1813176 and AST-2008108.
The Australia Telescope Compact Array is part of the Australia Telescope National Facility\footnote{\url{https://ror.org/05qajvd42}} which is funded by the Australian Government for operation as a National Facility managed by CSIRO. 
We acknowledge the Gomeroi people as the traditional owners of the Observatory site.
The ATCA data reported here were obtained under Programs C1303 (P.I.: P.~Lundqvist) and C1473 (P.I.: S.~D.~Ryder).
The SALT data reported here were taken as part of Rutgers University program 2021-1-MLT-007 (PI: S.~W.~Jha).
J.S.\ acknowledges support from the Packard Foundation.

\bibliography{zotero_abbrev}

\end{document}

%% file: affil.tex
\newcommand{\LCO}{\affiliation{Las Cumbres Observatory, 6740 Cortona Drive, Suite 102, Goleta, CA 93117-5575, USA}}
\newcommand{\UCSB}{\affiliation{Department of Physics, University of California, Santa Barbara, CA 93106-9530, USA}}

\newcommand{\UCD}{\affiliation{Department of Physics and Astronomy, University of California, Davis, 1 Shields Avenue, Davis, CA 95616-5270, USA}}

\newcommand{\OKC}{\affiliation{Oskar Klein Centre, Department of Astronomy, Stockholm University, Albanova University Centre, SE-106 91 Stockholm, Sweden}}

\newcommand{\UA}{\affiliation{Steward Observatory, University of Arizona, 933 North Cherry Avenue, Tucson, AZ 85721-0065, USA}}

\newcommand{\UNC}{\affiliation{Department of Physics and Astronomy, University of North Carolina, 120 East Cameron Avenue, Chapel Hill, NC 27599, USA}}

\newcommand{\GeminiNorth}{\affiliation{Gemini Observatory, 670 North A`ohoku Place, Hilo, HI 96720-2700, USA}}
\newcommand{\Keck}{\affiliation{W.~M.~Keck Observatory, 65-1120 M\=amalahoa Highway, Kamuela, HI 96743-8431, USA}}
\newcommand{\UW}{\affiliation{Department of Astronomy, University of Washington, 3910 15th Avenue NE, Seattle, WA 98195-0002, USA}}
\newcommand{\DiRAC}{\altaffiliation{DiRAC Fellow}}
\newcommand{\USask}{\affiliation{Department of Physics \& Engineering Physics, University of Saskatchewan, 116 Science Place, Saskatoon, SK S7N 5E2, Canada}}

\newcommand{\Rutgers}{\affiliation{Department of Physics and Astronomy, Rutgers, the State University of New Jersey,\\136 Frelinghuysen Road, Piscataway, NJ 08854-8019, USA}}

\newcommand{\Macquarie}{\affiliation{School of Mathematical and Physical Sciences, Macquarie University, NSW 2109, Australia}}
\newcommand{\AAARC}{\affiliation{Astronomy, Astrophysics and Astrophotonics Research Centre, Macquarie University, Sydney, NSW 2109, Australia}}

\newcommand{\MSU}{\affiliation{Center for Data Intensive and Time Domain Astronomy, Department of Physics and Astronomy,\\Michigan State University, East Lansing, MI 48824, USA}}

%% file: specplot_list.tex
\begin{deluxetable}{cccRl}
\tablecaption{Log of Spectroscopic Observations\label{tab:spec}}
\tablehead{\colhead{MJD} & \colhead{Telescope} & \colhead{Instrument} & \colhead{Phase}}
\startdata
59529.859 & SALT & RSS & -16.6 & \tablenotemark{a} \\
59530.868 & SALT & RSS & -15.6 \\
59531.118 & SOAR & Goodman & -15.3 \\
59531.266 & NTT & EFOSC & -15.2 \\
59531.703 & FTS & FLOYDS & -14.8 \\
59532.073 & SALT & RSS & -14.4 \\
59532.838 & SALT & RSS & -13.6 \\
59533.625 & FTS & FLOYDS & -12.9 \\
59533.856 & SALT & RSS & -12.6 \\
59534.498 & FTS & FLOYDS & -12.0 \\
59535.073 & SALT & RSS & -11.4 \\
59536.445 & FTS & FLOYDS & -10.0 \\
59536.855 & SALT & RSS & -9.6 \\
59538.070 & SALT & RSS & -8.4 \\
59538.853 & SALT & RSS & -7.6 \\
59550.598 & FTS & FLOYDS & +4.0 \\
59556.805 & SALT & RSS & +10.2 \\
59558.477 & FTS & FLOYDS & +11.9 \\
59559.087 & SOAR & Goodman & +12.5 \\
59560.992 & SALT & RSS & +14.4 \\
59561.478 & FTS & FLOYDS & +14.9 \\
59563.000 & SALT & HRS & +16.4 \\
59564.498 & FTS & FLOYDS & +17.9 \\
59567.672 & FTS & FLOYDS & +21.0 \\
59568.010 & SALT & RSS & +21.4 \\
59569.067 & SOAR & Goodman & +22.4 \\
59572.542 & FTS & FLOYDS & +25.9 \\
59577.480 & FTS & FLOYDS & +30.8 \\
59582.540 & FTS & FLOYDS & +35.8 \\
59589.594 & FTS & FLOYDS & +42.8 \\
59591.919 & SALT & HRS & +45.2 \\
59594.590 & FTS & FLOYDS & +47.8 \\
59597.890 & SALT & HRS & +51.1 \\
59615.419 & FTS & FLOYDS & +68.5 \\
59630.480 & FTS & FLOYDS & +83.5 \\
59634.105 & SOAR & Goodman & +87.1 \\
59648.447 & FTS & FLOYDS & +101.4 \\
59665.060 & SOAR & Goodman & +117.9 \\
59668.398 & FTS & FLOYDS & +121.2 \\
59681.390 & FTS & FLOYDS & +134.2
\enddata
\tablenotetext{a}{The spectrum presented here is a rereduction of the data published by \cite{bostroem_global_2021} on the Transient Name Server.}
\end{deluxetable}